\documentstyle[pre,aps,epsf]{revtex}
\input{psfig.tex}
\begin{document}

\draft

\title{Rational dynamical zeta functions for birational transformations}

\author{N. Abarenkova}
\address{Centre de Recherches sur les Tr\`es Basses Temp\'eratures,
B.P. 166, F-38042 Grenoble, France\\
Theoretical Physics Department, Sankt Petersburg State University,
Ulyanovskaya 1, 198904 Sankt Petersburg, Russia}

\author{J.-Ch. Angl\`es d'Auriac\footnote{e-mail
: dauriac@crtbt.polycnrs-gre.fr}}
\address{Centre de Recherches sur les Tr\`es Basses Temp\'eratures,
B.P. 166, F-38042 Grenoble, France}

\author{S. Boukraa}
\address{Institut d'A\'eronautique, Universit\'e de Blida, BP 270,
Blida, Algeria}

\author{S. Hassani\footnote{e-mail
: hassani@ist.cerist.dz}}
\address{CDTN, Boulevard F.Fanon, 16000 Alger, Algeria}

\author{J.-M. Maillard\footnote{e-mail
: maillard@lpthe.jussieu.fr}}
\address{ LPTHE, Tour 16, 1er \'etage, 4 Place Jussieu, 
75252 Paris Cedex, France}

\date{\today}

\maketitle

\begin{abstract}
We propose a conjecture for the exact expression of the 
dynamical zeta function
for a family of birational transformations of two variables,
depending on two parameters. This conjectured function is
a simple rational expression with integer coefficients.
This yields an algebraic value for the topological entropy.
Furthermore the generating function for the Arnold complexity
is also conjectured to be a rational expression with integer coefficients
with the same singularities as for the dynamical zeta function. 
This leads, at least
in this example, to an equality between the Arnold complexity
and the exponential of the topological entropy. We also give a semi-numerical method
to effectively compute the Arnold complexity.
\end{abstract}

\vskip .2cm 


\vskip .2cm


\section{Introduction}
To study the complexity of continuous, or discrete, dynamical
systems, a large number of concepts have been introduced
\cite{Ott,ASY}.
A non exhaustive list includes the Kolmogorov-Sinai
metric entropy~\cite{K58,S59}, the Adler-Konheim-McAndrew
topological entropy~\cite{AKM65}, the Arnold 
complexity~\cite{A}, the Lyapounov characteristic exponents, 
the various fractal dimensions, \cite{GP84,ER85}
the Feigenbaum's numbers of period-doubling 
cascades~\cite{F78,GreMcKViFeig81}, $ \, \cdots$ .
Many authors have tried to study and discuss the relations
between these various notions in an abstract 
framework ~\cite{HePro83,GraPro84}.
Inequalities have been shown, for instance the metric entropy
is bounded by the topological entropy, let us also mention
the Kaplan-Yorke relation~\cite{KY78,L81}. 
Furthermore, many specific dynamical systems have been
introduced enabling to see these notions at work.
Some of the most popular are the Lorentz system~\cite{L63},
the baker map~\cite{FY83}, the logistic map~\cite{M76},
the Henon map~\cite{H76}. Each of these systems
has been useful to understand and exemplify the previous
complexity measures.

Here, we introduce another two-parameter family of mapping
of two variables, originating from lattice statistical
mechanics for which much can be said. In particular, we will
conjecture an exact algebraic value for the  exponential of the
topological entropy
and Arnold complexity. Furthermore, these two measures of complexity are
found to be equal for all the values of the two
parameters, generic or not (the notion of genericity is
explained below). A fundamental distinction must be 
made between the previously mentioned complexity measures
according to their invariance under certain classes of transformations.
One should distinguish, at least, two different sets of complexity
measures, the ones which are invariant under the larger classes of 
variables transformations, like the {\em topological entropy}
or the {\em Arnold complexity}~\cite{A}, and 
the other measures of complexity
which also have invariance properties, but under a ``less larger''
set of transformations, and are therefore more sensitive to the
details of the mapping (for instance they will depend on the metric).

We now introduce the following two parameters family of birational
transformations $k_{\alpha,\epsilon}$:
\begin{eqnarray}
\label{uv}
u_{n+1} &=&\,  1 - u_n + u_n/v_n  \nonumber \\
v_{n+1} &=& \,\epsilon + v_n - v_n/u_n +\, \alpha \cdot (1 - u_n + u_n/v_n)
\end{eqnarray}
which can also be written projectively :

\begin{eqnarray}
\label{uvt}
u_{n+1} &=&\, (v_n t_n - u_n  v_n + u_n  t_n ) \cdot u_n \nonumber \\
v_{n+1} &=&\, \epsilon \cdot u_n \cdot  v_n \cdot  t_n 
+ (u_n \, - \, t_n) \cdot v_n^2  
+\alpha \cdot (v_n   t_n - u_n  v_n + u_n   t_n) \cdot u_n\nonumber \\
t_{n+1} &=&\,  u_n \cdot  v_n \cdot  t_n
\end{eqnarray}
As far as complexity
calculations are concerned,
the $\alpha=0$ case is singled out. In that case,
it is convenient to use a change of
 variables (see Appendix \ref{AppendixA}) to get the very
simple form $\, k_{\epsilon}$:
\begin{eqnarray}
\label{yz}
y_{n+1} \, &=& \, \, z_n +1 - \epsilon  \nonumber \\
z_{n+1} \, &=& \,\,   y_n \cdot \frac{z_n-\epsilon}{z_n + 1} 
\end{eqnarray}
or  on its homogeneous counterpart :
\begin{eqnarray}
\label{yzt}
y_{n+1} &=& \, (z_n + t_n - \epsilon \cdot  t_n) \cdot (z_n\, +t_n)\, ,
\nonumber  \\
z_{n+1} &=& \,  y_n \cdot (z_n - \epsilon  \cdot t_n)\, , \nonumber  \\
t_{n+1} &=&  \, t_n \cdot (z_n\,  + t_n)
\end{eqnarray}
These transformations derive from a transformation
acting on a $q \times q$ matrices $M$~\cite{BoMaRo94} :
\begin{equation}
\label{K}
K_q = t \circ I
\end{equation}
where $t$ permutes
the entries $M_{1,2}$ with $M_{3,2}$, and $I$ is the homogeneous
inverse: $I(M) = {\rm det}(M) \cdot M^{-1}$. Transformations of this
type, generated by the composition of permutations of the entries
and matrix inverse, naturally emerge in the analysis of lattice
statistical mechanics symmetries~\cite{BoMaRo95}.


\section{The complexity growth}
The correspondence~\cite{BoMaRo94} between
transformations $K_q$ and $k_{\alpha,\epsilon}$,
more specifically between  $K^2_q$ and $k_{\alpha,\epsilon}$,
is given in Appendix \ref{AppendixA}. It will be shown below
that, beyond this correspondence, $K^2_q$ and $k_{\alpha,\epsilon}$
share properties concerning the complexity. Transformation $K_q$ is
homogeneous and of degree $(q-1)$ in the $q^2$ homogeneous entries.
When performing the $n^{\rm th}$  iterate one expects a growth of the
degree of each entries as $(q-1)^n$. It turns out  that, at each step of the
iteration, some factorization of all the entries occurs. The common factor 
can be factorized out in each entry leading to a reduced matrix $M_n$, which
is taken as the representent of the $n^{\rm th}$ 
iterate in the projective space.
Due to these factorizations the growth of the calculation is not $(q-1)^n$
but rather $\lambda^n$ where generically
$\lambda$ is the largest
root of $1 + \lambda^2 - \lambda^3 = 0$ 
(i.e. 1.46557123 $< q-1$)~\cite{BoMaRo94,BoMaRo93c}.
We call $\lambda$ the complexity growth or simply the complexity.
This result is a consequence of a {\em stable factorization
scheme}\footnote{Complexity
growth can also be understood from a singularity
 point of view~\cite{FaVi93}, or through
recurrence relations associated with the geometry
of the singularities of the 
evolution~\cite{BeVi98}.  This is not the approach developped
here.} given in
Appendix \ref{AppendixB}, from which two generating functions
$\alpha(x)$ and $\beta(x)$ can be constructed. Generating function $\alpha(x)$
keeps track respectively of the degrees of the determinants of the
successive reduced matrices
and $\beta(x)$ of the degrees of the successive common factors.
The function $\alpha(x)$ should not be confused
with the parameter $\alpha$.
The actual value of $\lambda$ is the inverse of the pole of
$\beta(x)$ (or $\alpha(x)$)
of smallest modulus. The algebraicity of the complexity is, 
in fact, a straight
consequence of the rationality of functions $\alpha(x)$ and $\beta(x)$ with
integer coefficients~\cite{BoMaRo94}. 
The same calculations have also been performed
on transformations Eq.~(\ref{uv}) and Eq.~(\ref{uvt}). 
In that case, factorizations
also occur at each step, and generating functions can be calculated.
These generating functions are, of course, different from the generating
functions for $\, K^2_q$ (see~\cite{BoMaRo94}) 
but they have the {\em same} poles, and 
consequently the same complexity growth. 
One sees that, remarkably, the complexity
$\lambda$ does not depend on the birational representation considered:
$K^2_q$ for any value of $q$, $k_{\alpha,\epsilon}$ or the 
homogeneous transformation Eq.~(\ref{uvt}).
It will be usefull to define some degree generating functions $\, G(x)$ :
\begin{equation}
\label{gendeg}
G(x) \, = \; \sum_{n}{d_n \cdot x^n}
\end{equation}
where $\, d_n$ is the 
degree of some quantities we look at, at each iteration step
(numerators or denominators of the two components
of $\, k^n$, degree of the entries 
of the ``reduced'' matrices $M_n$'s, degree of
the extracted polynomials 
$\, f_n$'s in Appendix \ref{AppendixB} ...). 
The complexity growth $\lambda$ is the inverse 
of the pole of smallest modulus of any of these degree generating functions 
$G(x)$ :
\begin{equation}
\label{lambda}
\log{\lambda} = \lim_{m \rightarrow \infty}{\frac{\log{ d_m }}{m}}
\end{equation}

\subsection{Complexity growth for $\, \alpha \, = \, 0$}
In the $\, \alpha \, = \, 0\, $ case, which corresponds 
to a codimension one variety of the parameter
space (see Appendix \ref{AppendixA} and \ref{AppendixC}),
 additional factorizations occur
reducing further the growth of the complexity. 
The generating functions are modified
and the new complexity is given, for $K_q$,  by the equation: 
\begin{equation}
\label{complex1}
1 - \lambda^2 - \lambda^4\,  = \, \, 0
\end{equation} 
i.e.  $\lambda \, \simeq \, 1.27202 \cdots$ .
For $k_\epsilon$, which corresponds to $K^2_q$,  the equation reads:
\begin{equation}
\label{complex}
1 - \lambda - \lambda^2\,  = \, \, 0
\end{equation} 
leading to the complexity  $\, \lambda \, \simeq \, 1.61803 \cdots  \,
\simeq \,(1.27202 \cdots)^2$.
Not surprisingly, the complexity of the mappings $\,
k_{\alpha,\epsilon}$ for $\alpha=0$ (see (\ref{uv})) 
and mapping $k_\epsilon$ (see (\ref{yz})), are the same: 
complexity $\lambda$ corresponds to the
asymptotic behavior of the degree of the successive quantities encountered
in the iteration (see (\ref{lambda})). Clearly, 
this behavior remains unchanged under simple
changes of variables. Note that this complexity growth\footnote{Growth
of the calculations related with factorizations were also
introduced by Veselov for some particular Cremona 
transformations~\cite{Ve89,MoVe91,Ve92}.}
 analysis can be performed
directly on  transformation  $\, k_\epsilon$, or on its homogeneous
counterpart Eq.(~\ref{yzt}).
The number of generating functions in the two cases is not the same, but 
all these functions lead to the same complexity. In fact, complexity $\lambda$
is nothing but the Arnold complexity~\cite{A}, known to be invariant
under transformations corresponding to a change of
variables (like the change of variables from  Eq.~(\ref{uv}) (for $\alpha=0$)
to Eq.~(\ref{yz}) or to Eq.~(\ref{yzt})).
Let us also recall that the Arnold complexity counts the number
of intersection between a fixed line\footnote{Or the 
intersection of the $n$-th iterate of
 any fixed algebraic curve together with any other possibly
different but fixed  algebraic curve.}
and its $n^{\rm th}$ iterate,
which clearly goes as $\lambda^n$. Conversely, all these calculations
can be seen as a handy way of calculating the Arnold complexity.

\begin{figure*}
\psfig{file=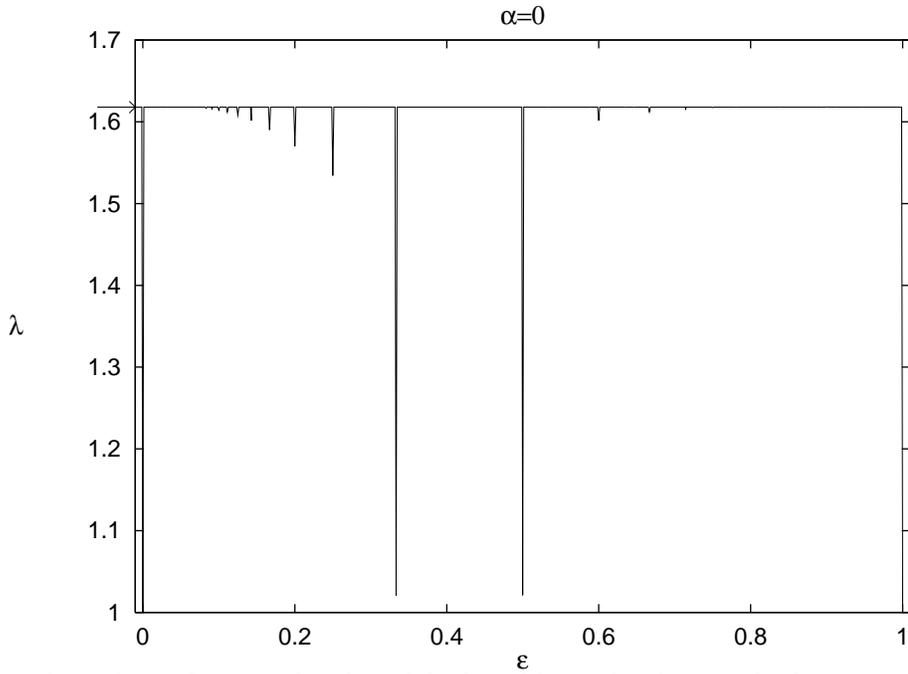}
\caption{ Complexity $\lambda$ as a function of $\epsilon$ taken of the form
$M/720$ plus the special values 1/7, 1/11, 1/13 and 5/7
for $\alpha=0$.
The arrow indicates the expected value.
\label{f:fig1}
}
\end{figure*}

All these considerations allow us to design a semi-numerical method to get
the value of the complexity growth $\lambda$ for any value
of the parameter $\epsilon$. The idea is to iterate, with (\ref{yz})
(or (\ref{uv})), a generic
{\em rational} initial point $(y_0,z_0)$ and to follow the magnitude
of the successive numerators and denominators.
During the first few steps, some accidental simplifications may occur,
but, after this transient regime, the integer denominators (for instance) grow
like $\lambda^n$ where $n$ is the number of iterations.
Typically, a best fit of the logarithm of the numerator as a linear function
of $n$, between $n=10$ and $n=20$, gives the value of $\lambda$ within
an accuracy of $0.1\%$. An integrable mapping
corresponds to a polynomial growth of the calculations :
the value of the complexity $\lambda$  has to be numerically
very close to $\, 1\, $.
Fig.~\ref{f:fig1} shows the values of the complexity
as a function of the parameter $\epsilon$. The calculations have
been performed using an infinite-precision C-library~\cite{C}.

For most of the values of $\epsilon$ we have found 
$\lambda \, \simeq\, 1.618$, 
in excellent agreement with the value predicted in Eq.~(\ref{complex}).
In~\cite{BoHaMa97}, it has been shown that the simple
rational values  $\epsilon=-1, 0, 1/3, 1/2, 1$ yield integrable 
mappings. For these special values one gets $\lambda \approx 1$
corresponding to a {\em polynomial} 
growth~\cite{BoHaMa97}. In addition, Fig.~\ref{f:fig1}
singles out two sets of values $\{ 1/4,\,  1/5, \, 1/6, \cdots ,\,  1/13 \}$
and $\{ 3/5,\, 2/3\, , \, 5/7 \}$, suggesting two infinite sequences
$\epsilon = 1/n$  and $\epsilon = (m-1)/(m+3)$\footnote{Note that 
$ \, m \, \, \rightarrow (m+3)/(m-1)\, $ is an involution.}
for $n$ and $m$ integers such that $n \ge 4$ and
$m \ge 7$ and $m$ odd. 
We call ``specific'' the values of $\epsilon$ of one of
the two forms above (together with
 the integrable values), and ``generic'' the others.
To confirm this conjecture, we go back to 
(the matrix) transformation $\, K_q$, for $q=3$, to get a generating function
of the degrees of some factors (the $f_n$'s in 
Appendix \ref{AppendixB}) extracted at each step
of iteration, namely, with the notations
of~\cite{BoMa95,BoMaRo93c,BoMaRo95} and of Appendix \ref{AppendixB},
function $\beta(x)$. From now on, we will give
 below, instead of $\beta(x)$, the expression of the following 
complexity 
generating function defined, for
 $\, q \times q\, $ matrices, as :
\begin{equation}
G_\epsilon^\alpha(q,\, x) \, = \; \, {{\beta(x)} \over {q \cdot x}}
\end{equation}
In the following the calculations are often displayed for $\, 3 \times 3$
matrices
and $\, G_\epsilon^\alpha(q,\, x) \,$ will simply be denoted 
 $\, G_\epsilon^\alpha( x) \,$.
Let us recall that the value of the complexity $\, \lambda$ 
is the inverse of
the root of smallest modulus of the denominator of this 
rational function. Examples of these calculations
in order to get the corresponding factorization
scheme and deduce the generating function $\beta(x)$ 
or $\, G_\epsilon^\alpha(x) \,$,
are given in  Appendix 
\ref{AppendixB}. In Appendix 
\ref{AppendixC}, we show how to
choose an initial matrix to iterate, the matrix 
satisfying $\alpha=0$ and $\epsilon = p/q$
for any integers $p$ and $q$. First, we have obtained
(see Appendix \ref{AppendixB})
 the generating function $G_\epsilon(x)$
in the generic case for $\alpha=0$ :
\begin{equation}
\label{first}
 G_\epsilon(x) \, =   \, \,   \, {{ 1+x+x^3} \over {1-x^2-x^4}}
\end{equation}
 We also got the generating function $G_\epsilon(x)$
for the different ``specific'' cases :
\begin{equation}
\label{betabeta1}
G_{1/m}(x)\,  =  \, \,  \,  {{1+x+x^3-x^{2 m+1}-x^{2 m +3}} \over
                         {1-x^2-x^4+x^{2 m + 4}}}\, , \qquad
\quad \hbox{with} \quad \, m \, \ge \, 4  
\end{equation}
\begin{equation}
\label{betabeta2}
G_{(m-1)/(m+3)}(x)\, = \, \,  \, {{1+x+x^3-x^{2 m + 6}} \over
			  {1-x^2-x^4+x^{2 m + 4}}} \, ,\qquad \quad  
\hbox{with}
                         \quad \, m \, \ge \, 7  \, 
                         \quad \, m \quad  \hbox{odd} 
\end{equation}
and :
\begin{eqnarray}
\label{betabeta3}
G_{\rm int}(x)\,  &=& \, {{ 1+x+x^3+x^4+x^8+x^{12}}   \over  
            {1-x^2-x^6+x^8-x^{10}+x^{12}+x^{16}-x^{18} }} \\
  &=&  \, {\frac {1+x\cdot (1\, +x^{2} )+{x}^{4}\cdot (1+x^4\,
+x^8 )}{1-x^2 \cdot (1\, -x^{12} )\, -\, x^6 \cdot
(1-x^{2}+x^4-x^6+{x}^8-x^{10} +{x}^{12})   }} \nonumber 
\end{eqnarray}
for the two integrable values $\epsilon = 1/2$ and $\epsilon = 1/3$.
For $\epsilon=1/m$ ($\, m \ge \, 4$) and
$\epsilon=(m-1)/(m+3)$ ($\, m \ge \, 7$ and $m$ odd),
the corresponding complexities are the inverse of the roots of smallest
modulus of polynomial :
\begin{equation}
\label{complexspec}
1 -x^2 -x^4 - x^{2\, m + 4}\, \,   = \,  \, \, 0
\end{equation}
 in agreement with the values of
Fig.~\ref{f:fig1}. In this figure the $\epsilon$-axis has been
discretized as $\, M/720$ ($M\, $ integer) 
and the extra values 1/7, 1/11, 1/13 and 5/7  have been added. 
This semi-numerical method acts as an `integrability detector'
and, further, provides a simple and efficient way to determine the complexity
of an algebraic mapping. Applied to mappings 
Eq.~(\ref{uv}), Eq.~(\ref{K}),
or Eq.~(\ref{yz}), it shows that the complexity 
is, generically, {\em independent} of the
value of the parameter $\epsilon$, except for the four integrable points,
and for two denombrable sets of points\footnote{
These two sets of points also appear naturally 
in the framework of a ``singularity
confinement analysis''~\cite{RaGra}.}. 

It is worth noticing that these results are not specific to 
$3 \times 3$ matrices, for example relation  Eq.~(\ref{first}) 
is  actually valid simply replacing
 $\, G_\epsilon^\alpha( x) \,$ by $\, G_\epsilon^\alpha(q,\, x) \,$.

\begin{figure}
\psfig{file=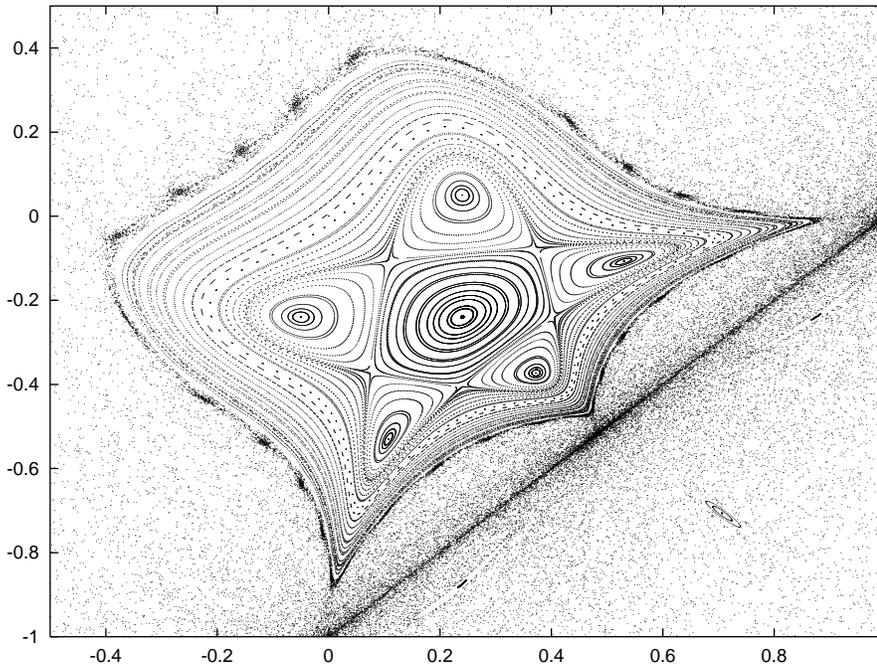}
\caption{Phase portrait of $k_\epsilon$ for $\alpha \, = \, 0\, $ and 
$\epsilon=13/25$. 550 orbits of length 1000 have been generated.
50 orbits start from points randomly choosen near a fixed point
of order 5 of $k_{\epsilon}\, =\, k_{13/25}$, and 500 others orbits
start from randomly choosen points outside the elliptic region.
Only the points inside the frame are shown.
\label{f:fig3}
}
\end{figure}


\subsection{Complexity growth for $\alpha \ne 0$}
\label{secB}
\begin{figure}
\psfig{file=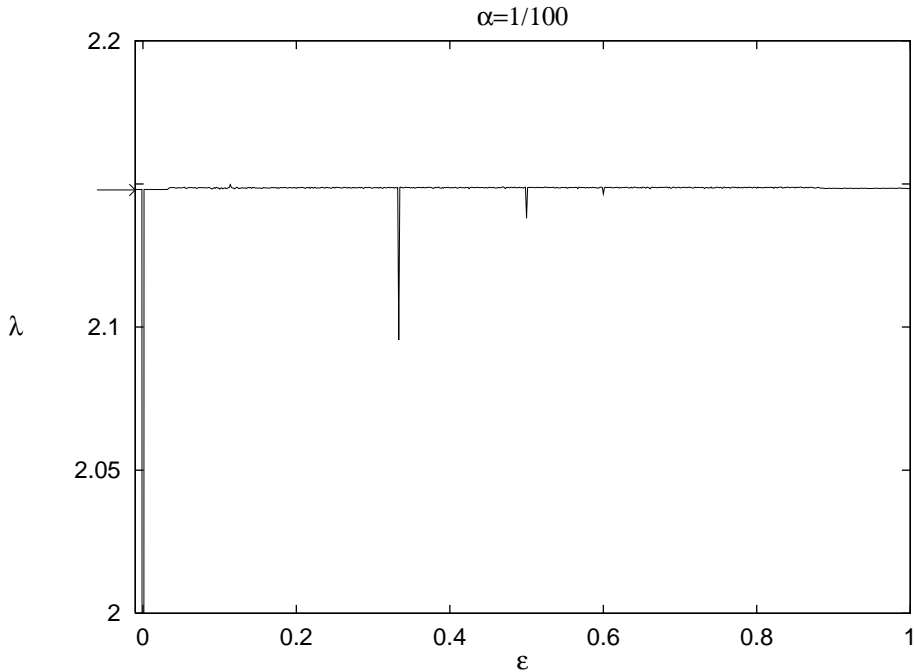}
\caption{Complexity $\lambda$ as a function of $\epsilon$ taken of the form
$M/720$ for $\alpha = 1/100$.
The arrow indicates the expected value.
\label{f:fig2}
}
\end{figure}

These complexity growth calculations can 
straightforwardly be generalized to  $\alpha \ne 0$.
As explained in Appendix \ref{AppendixB}, 
the ``generic'' generating function is :
\begin{eqnarray}
\label{alpeps}
G_\epsilon^\alpha(x)  \, = \, \, 
{{ 1+x^2 } \over { 1\, - \, x\, -\, x^3 \, }}
\end{eqnarray}
The pole of smallest modulus of Eq.~(\ref{alpeps}) gives
$1.46557 \cdots $ for the value of the complexity 
for the matrix transformation $K$. The complexity
for the transformation $k_{\alpha,\epsilon}$ is 
the square of this value: $\lambda \, = \, 2.14790 \cdots$ .
Fig.~\ref{f:fig2} shows, for $\, \alpha \, = \, 1/100 $,
complexity $\lambda$  as a function 
of the parameter $\, \epsilon\, $, obtained
with the semi-numerical method previously explained.
Even with such a ``small value'' of $\, \alpha \, $ the expected drastic 
change of value of the complexity (namely 
$1.61803 \,\rightarrow \, 2.14790$)
is non-ambiguously seen. 
Moreover, Fig.~\ref{f:fig2} clearly shows
that, besides the  value $\, \epsilon \, = \,0$ 
 known to be integrable whatever $\alpha$
\cite{BoHaMa97}, at least the following
 values $\epsilon=1/2$, $\epsilon=1/3$  and  $\epsilon=3/5$
are associated with a significantly smaller complexity,
at least for the discretization in $\epsilon$ we have investigated.
From these numerical results and by analogy 
with $\alpha =0$, one could figure out that all the 
$\epsilon \, = \, 1/m$  are also  non-generic 
values of $\epsilon$. In fact a factorization
scheme analysis like the one depicted 
in Appendix \ref{AppendixB}) shows that  $\epsilon \, = \, 1/4$
or $\epsilon \, = \, 1/7$ actually
correspond to the generic generating function Eq.~(\ref{alpeps}).
We got similar results for other values of  $\alpha \ne 0$.
However, when varying $\alpha$ and keeping $\epsilon$ fixed,
new values of the complexity $\lambda$ occur, $\lambda$ being 
some ``stair-case'' function of $\alpha$. 
We will not exhaustively describe the rather involved
``stratified'' space in the $(\alpha\, , \, \epsilon)\, $ 
plane, corresponding to the various ``non generic''
complexities. Let us just keep in mind that,
besides  $\, \epsilon \, = \,0$ and  $\, \epsilon \, = \,-1$,
at least $\, \epsilon \, = \, 1/2\, $, $\, \epsilon \, = \, 1/3\, $
and $\, \epsilon \, = \, 3/5\, $ are singled out for $\alpha \ne 0$
in our semi-numerical analyzis. 
The generic expression (for $3 \times 3$ matrices) for
the generating function  $ \,G(x)\, \, $ Eq.~(\ref{alpeps})
is replaced,  for the ``non-generic'' $\, \epsilon \, = \,1/2$ 
  (with $\alpha \ne 0$),  
by :
\begin{equation}
\label{alpeps1sur2}
G_{1/2}^\alpha(x) \, =  \, 
\frac {1\, +x\, +{x}^{3}\, -{x}^{16}}
      {(1-\, x^2 )\cdot 
        (1\, -{x}^{2}\, -{x}^{4}\, -2\,{x}^{6}\, -{x}^{8}\, -2\,{x}^{10}\,
	-{x}^{12}\, -{x}^{14} )}
\end{equation}
For the other ``non-generic'' value
of $\epsilon$, $\, \epsilon \, = \,1/3$,  the 
complexity generating function 
reads :
\begin{equation}
\label{alpeps1sur3}
G_{1/3}^\alpha(x) \, = \,\frac { 1+x+{x}^{3}\, -{x}^{12}}
{(1-x^2 ) \cdot (1-{x}^{2}-{x}^{4}-2\,{x}^{6}-{x}^{8}-{x}^{10})}
\end{equation}
For  the  ``non-generic'' value
 $\, \epsilon \, = \,3/5$,  the 
complexity generating function reads :
\begin{equation}
\label{g35}
G_{3/5}^\alpha(x) \, =  \, \frac {1+x+{x}^{3}-{x}^{20}}
{(1-x^2) \cdot (1-x^2-x^4-2 x^6 -x^8-2 x^{10}- x^{12}-2 x^{14}-x^{16}-x^{18})}
\end{equation}

\section{Dynamical zeta function and  topological entropy}
It is well known that the fixed points of the successive powers
of a mapping are extremely important in order to understand 
the complexity of the phase
space. A lot of work has been devoted to study these fixed
points (elliptic or saddle fixed points, 
attractors, basin of attraction, etc),
and to analyse related concepts (stable and unstable manifolds, 
homoclinic points, etc). 
We will here follow another point of view and study the generating
function of the {\em number} of fixed points. 
By analogy with the Riemann $\zeta$ function,
Artin and Mazur~\cite{AM65} introduced
a powerful object the so-called {\em dynamical zeta function}:
\begin{equation}
\label{zeta}
\zeta(t)\,  =\, \,  \exp{ \left( \sum_{m=1}^{\infty}{\# {\rm fix}(k^m)}\cdot
 \frac{t^m}{m} \right) }
\end{equation}
where $\#{\rm fix}(k^m)$ denotes the number of fixed points of $\, k^m$.
The generating functions 
\begin{equation}
\label{genptfixe}
H(t)=\sum{\#{\rm fix}(k^m)} \cdot  t^m 
\end{equation}
can be deduced from the $\zeta$ function:
\begin{equation}
\label{Gfromzeta}
H(t)\, = \, \, \, t \frac{\rm d}{\rm dt}(\log{\zeta(t)}).
\end{equation}
The topological entropy $h$ is related to the singularity
of the dynamical $\zeta$ function:
\begin{equation}
\label{h}
\log{h} = \lim_{m \rightarrow \infty}{\frac{\log{( \# {\rm fix}(k^m))}}{m}}
\end{equation}
If the dynamical zeta function can be interpreted as the ratio
of two characteristic polynomials of two linear operators\footnote{For
more details on these Perron-Frobenius, or Ruelle-Araki transfer
operators, and other shifts on Markov's partition in a symbolic
dynamics
framework, see for instance~\cite{Bo73,Bo75,Ru78,Ru91}.}
$A$ and $B$,
namely $\zeta(t) = {\rm det}(1\, -\, t \cdot B) 
/ {\rm det}(1\, -\, t \cdot A) $,
then the number of fixed points  $\#{\rm fix}(k^m)$ can be expressed from
${\rm Tr}(A^n)-{\rm Tr}(B^n)$. 
In this linear operators framework, the {\em rationality} of 
the $\zeta$ function, and therefore the algebraicity
of the  exponential of the topological entropy, 
amounts to having a {\em finite dimensional
representation} of the linear operators $\, A$ and $B$.
In the case of a rational $\zeta$ function, the  exponential of the
topological
entropy is the inverse of the pole of smallest modulus.
Since the number of invariant points remains unchanged under
topological conjugaison (see Smale \cite{S67} for this notion),
the $\zeta$ function is also a topologically
invariant function, invariant under a large set of transformations,
and does not depend on a specific choice of variables.
Such  invariances were also noticed for the complexity growth
$\lambda$. It is then tempting to make a connection between
the {\em rationality} of the complexity generating 
function previously given, 
 and a possible {\em rationality} of the dynamical
$\zeta$ function. We will also compare the Arnold complexity $\lambda$
and the  exponential of the topological entropy $h$.

\subsection{Dynamical zeta function for $\alpha=0$, $\epsilon$ generic}
We try here to get the expansion of the dynamical zeta
 function of the mapping
$k_\epsilon$ (see Eq.~(\ref{yz})), for generic values of $\epsilon$
which are neither of the form $1/m$, nor of the form $(m-1)/(m+3)$.
We concentrate on the value $\epsilon\, =\, 13/25\, =\, 0.52$. This value is
close to the value 1/2 where the mapping is integrable~\cite{BoHaMa97}. On can
gain an idea of the number, and localization, of the (real) fixed points
looking at the phase portrait of Fig.~\ref{f:fig3}. The elliptic fixed
point $(y_0,z_0)=\, (.24\, ,\, -.24)\, $ is well seen, as well as the five
elliptic points and the five saddle points of $k_{\epsilon}^5$.
Many points of higher degree are also seen.
Transformation $k_\epsilon$ has a single fixed point for any 
$\epsilon$. This fixed point is elliptic for $\epsilon \ge 0$
 and localized at
$(y_0,z_0)=((1-\epsilon)/2,\, (\epsilon-1)/2)$. Transformation
$k_\epsilon^2$ has only the fixed point inherited from $\, k_\epsilon$.
The new fixed points of $k_\epsilon^3$ are 
$( 2\, - \, \epsilon\, , (\epsilon-1)/2 )$, $(-1, 1)$ 
and $((1-\epsilon)/2, \epsilon - 2)$.
Transformation $k_\epsilon^4$ has four 
new fixed points. At this point the calculation
are a bit too large to be carried out with a literal $\epsilon$,
and we particularize $\epsilon\, =\, 13/25$. For $k_\epsilon^5$
we have five new elliptic points and five new saddles points.
The coordinates $z$ and $y$ of these points are roots 
of the two polynomials:
\begin{eqnarray}
\label{fixedptsk5_1}
P(z)\, &=& \, 
(4375\,z^2+1550\,z-89 )\, (175\,z^2\, +106\,z+7 ) \,  (25\,z^2\,+12\,z\, +1 )^2
\left (25\,z\, +6\right )^3 \\
Q(y) \, &=& \, P(-y) \label{fixedptsk5_2}
\end{eqnarray}
        
The five pairings of the seven roots of Eq.~(\ref{fixedptsk5_1}) and
Eq.~(\ref{fixedptsk5_2}), giving the
five elliptic points, are 
(0.530283,\, -0.107335),  $\, \, \, $
(-0.050283,\, -0.24),
(0.372665,\, -0.372665), 
(0.107335,\,-0.530283),
(0.24,\, 0.050283) and
the five pairings giving the five hyperbolic-saddle points are   
(0.372665,\, -0.075431), 
(0.107335,\, -0.107335),
(0.404568,\, -0.24), 
(0.075431,\, -0.372665,), 
(0.24,\, -0.404568).
This is clearly seen on  Fig.~\ref{f:fig3} where the occurrence of 
five ``petals'' corresponding to five elliptic points are obvious,
the  five hyperbolic points being located between the petals.

For transformation $\, k_{\epsilon}^6\, $, beyond the fixed points of $k$ and
$k^3$, one gets two complex saddle fixed points, i.e. transformation $k$
has two 6-cycles.
For transformation $\, k_{\epsilon}^7\, $, one obtains one elliptic real
fixed point, one saddle real fixed point and and two complex
saddle fixed points.
For transformation $\, k_{\epsilon}^8\, $, one obtains one saddle real
fixed point and four complex saddle fixed points.
For transformation $\, k_{\epsilon}^9\, $, one obtains one elliptic real
fixed point, three saddle real fixed points and and four complex
saddle fixed points.
For transformation $\, k_{\epsilon}^{10}\, $, one obtains one elliptic real
fixed point, one saddle real fixed point and and three complex
elliptic fixed points and six saddle complex fixed points.
The two elliptic fixed points of $k_{\epsilon}^{10}$ (0.24,\, -0.874) and
(0.874,\, -0.24) are seen as ``ellipse'' on Fig.~(\ref{f:fig3}).
For transformation $\, k_{\epsilon}^{11}\, $, one obtains one elliptic real
fixed point, five saddle real fixed point and and twelve complex
saddle fixed points. On Fig.~(\ref{f:fig3}) a fixed point
of $k_{\epsilon}^{12}$ lying on $y+z=0$ is seen near $y=\, -13/25$.
The polynomials, similar to Eq.~(\ref{fixedptsk5_1}) and 
Eq.~(\ref{fixedptsk5_2})
(or to Eq.~(\ref{P}) given in appendix \ref{AppendixD}), as well as the 
specific pairing of roots, for the successive iterates $k^N$, are available
in~\cite{ftpano}.

It is worth noticing, that among the 53 cycles of $k_\epsilon$
of length smaller, or equal, to 11, as much as 44 have a representent
 on the line
$y+z=0$, six have one on the line $\,y\,+\bar{z}\,=\, 0$. Two of 
the three remaining cycles are of length 11, while the last 
is of length eight.
The particular role played by the  $\, y\, + \, z \,=\,  0\, $
line can be simply understood. Let us calculate the inverse 
of the birational transformation  
(\ref{yz}). It has a very simple form :
\begin{eqnarray}
&&z_{n+1} \, = \, y_n  -(1 - \epsilon)\, ,  \qquad 
y_{n+1}  \, = \, z_n \cdot 
{{ y_n \, + \epsilon} \over { y_n \, -1}}
\end{eqnarray}
which is nothing but transformation (\ref{yz})
where $\, y_n\, $ and $\, -\, z_n$ have been permuted.
The  $\, y_n\,\,\leftrightarrow \,\, -\, z_n$ symmetry just corresponds to 
the {\em time-reversal symmetry }  $\, k_{\epsilon} 
\,\,\leftrightarrow
\,\, k^{-1}_{\epsilon}\, $ transformation.
The  $\, y\, + \, z \,=\,  0\, $
line is the {\em  time-reversal invariant line}.
Also note that only one of the 31 complex cycles is of
the form $Z_0,Z_1, \cdots Z_p, \bar{Z_0},\bar{Z_1}, \cdots \bar{Z_p}$
where $Z_i=(y_i,z_i)$ and $\bar{Z}_i$ is the complex conjugate. 
The 30 remaining complex cycles are actually 15 cycles and their
complex conjugates. 

Eventually, we observe an  
 {\em area preserving}~\cite{GiLaSi97} property  {\em in the neighborhood
of all the fixed points of} $k_{\epsilon}^n$ : 
the product of the modulus of the two
eigenvalues of the Jacobian (i.e. the determinant) of $k_{\epsilon}^n$, at all 
fixed points for $ n \le 11$, 
is equal to $\, 1\, $. This {\em local} property is rather non trivial :
the determinant of the product of the jacobian over an {\em incomplete}
cycle is very complicated and only when one multiply by the
last jacobian does the product of the determinants shrink to $\, 1\,$.

The total number of fixed points of $\,k_{\epsilon}^N$
for $\,N\, $ running from $1$ to $\, 11$, yields 
the following expansion, up to order eleven, for 
the generating function $H(t)$ of the  number of fixed points:
\begin{equation}
\label{G}
H_\epsilon(t) \, = \,\,\, t+{t}^{2}+4\,{t}^{3}+5\,{t}^{4}\,
+11\,{t}^{5}+16\,{t}^{6}+29\,{t}^{7}+45\,{t}^{8}+76\,{t}^{9}
 +121\,{t}^{10}+199\,{t}^{11}\,
\, + \cdots
\end{equation}
This expansion coincides with the one of the {\em rational}
function :
\begin{eqnarray}
\label{GG}
H_\epsilon(t) \, = \,  \, 
{\frac {t \cdot \left (1 \, + \, t^2 \right )}
{\left (1-t^2\right )\left (1-t-t^2\right )}}
\end{eqnarray}
which corresponds to a very  {\em simple rational} expression for
the dynamical zeta function:
\begin{eqnarray}
\label{conjec}
\zeta_\epsilon(t)  \, \,  = \, \,  \,  {{1\, -t^2 } \over {1\, -t\, -t^2}} 
\end{eqnarray}
Expansion (\ref{G}) remains unchanged for all the other
generic  values of $\epsilon\, $ we have also studied.

We conjecture that :
 {\em The  simple rational expression}
Eq.~(\ref{conjec}) {\em is the actual  expression 
of the dynamical zeta function for any generic value of $\epsilon$}.

Comparing the expression Eq.~(\ref{complex}) with  Eq.~(\ref{conjec}),
one sees that the singularities of the dynamical zeta function
happen to coincide with the singularities of the 
generating functions of the Arnold complexity.
In particular, the complexity growth $\lambda$ 
and the exponential of the topological entropy $h$ are {\em equal}.

When mentioning zeta functions, it is tempting to seek 
for simple {\em functional relations} relating $\zeta(t)$ and 
 $\zeta(1/t)$. Let us introduce the following ``avatar'' of the
dynamical  zeta function :
\begin{eqnarray}
\label{avatar}
\widehat{\zeta}(t)  \, \,\,=\, \,
\,\, {{ \zeta(t)} \over { \zeta(t) \, -\, 1}}
\end{eqnarray}
The transformation $ z \, \, \rightarrow \, \, z/(z-1)\, $ 
is an involution.
One immediately verifies that $\, \widehat{\zeta_\epsilon}(t)\, $
corresponding to (\ref{conjec})  verifies two extremely
simple and remarkable functional relations :
\begin{eqnarray}
\label{Waouh1}
\widehat{\zeta}_\epsilon(t)\,  \, \, 
= \,\,\, \,- \,  \widehat{\zeta}_\epsilon(1/t)\, , \quad
\qquad \hbox{and :} \qquad  \quad\widehat{\zeta}_\epsilon(t)\,  \, \, 
= \,\,\, \,\,  
\widehat{\zeta}_\epsilon(-1/t)\, ,
\end{eqnarray}
or on the zeta function $\, \zeta(t)\, $ :
\begin{eqnarray}
\zeta_\epsilon(1/t) \, = 
 \, \, {{ \zeta_\epsilon(t) } \over 
{ 2\cdot \zeta_\epsilon(t) \, - \, 1}} \, , \quad
\qquad \, \hbox{and :} \qquad \quad \zeta_\epsilon(-1/t) \, = 
\, \, \zeta_\epsilon(t)
\end{eqnarray}
The generating function (\ref{GG}) verifies :
\begin{eqnarray}
\label{gg}
H_\epsilon(-1/t)\, \,= \,\,\, -\, H_\epsilon(t)
\end{eqnarray}
An alternative way of writing the dynamical zeta functions
 relies on the decomposition of the 
fixed points into {\em cycles} which corresponds 
to the Weyl conjectures~\cite{W}. Let us introduce $\, N_r\, $ the number
of irreducible cycles of 
$\, k_{\epsilon}^r$: for instance for  $\, N_{12}\, $
we count the number of fixed points of $k_|epsilon^{12}$, that are not fixed
points of  $\, k_{\epsilon}$, $\, k_{\epsilon}^{3}$,
 $\, k_{\epsilon}^{4}$ or $\, k_{\epsilon}^{6}$, and divide by twelve.
One can write the dynamical zeta function as :
\begin{eqnarray}
\label{Weyl}
\zeta_\epsilon(t) \, = \,\,\, {{1} \over {(1-t)^{N_1}}} \cdot 
{{1} \over {(1-t^2)^{N_2}}} \cdot
 {{1} \over {(1-t^3)^{N_3}}} \,\cdots\,
 {{1} \over {(1-t^r)^{N_r}}}\, \cdots\,
\end{eqnarray}
The combination of the $\, N_r$'s, inherited from 
the product (\ref{Weyl}), automatically takes into account
the fact that the total number of fixed points of  $k_{\epsilon}^{r}$
can be obtained from fixed points of  $k_{\epsilon}^{p}$, where $\, p$
divides $r$, and from irreducible  fixed points of  $\, k_\epsilon^{r}$
itself (see~\cite{W} for more details).
A detailed analysis of this cycle decomposition (\ref{Weyl})
for generic values of $\, \epsilon\, $ 
will be detailed elsewhere~\cite{BoHaMa98}.
The previous exhaustive list of fixed points (up to order twelve)
can be revisited in this irreducible cycle decomposition
point of view. The results of~\cite{ftpano}  yield :
$N_1\, = \, 1\, , \, N_2\, = \, 0\,, \,  N_3\, = 1\, , 
\,  N_4\, = 1\, ,\,  N_5\, = 2\, ,  \,  N_6\, = 2\, , 
\,  N_7\, = 4\, , \,  N_8\, = 5\, ,$$ \,  N_9\, = 8\, , 
\,  N_{10}\, = 11\, , \,  N_{11}\, = 18\, \,$.
One actually verifies easily that (\ref{conjec})
and (\ref{Weyl}) have the same expansion up to order twelve
with these values of the  $\, N_r$'s. 
The next  $\, N_r$'s should be  $\,  N_{12}\, = 25\,, \,  N_{13}\, = 40\, , 
\,  N_{14}\, = 58\, , \,$
$  N_{15}\, = 90\, , \, \cdots $

It should be noticed that if one introduces some generating function
for the {\em real} fixed points of $k^N$, this generating function
has the following expansion, up to order eleven, for $\epsilon \, = \,
.52\,$ :
\begin{eqnarray}
\label{reel}
H^{\rm real}_{\epsilon} \, = \,  t+{t}^{2}+4\,{t}^{3}+5\,{t}^{4}+11\,{t}^{5}
+4 \, {t}^{6}
+15 \,{t}^{7}+
13 \,{t}^{8}+40\,{t}^{9}
 +31\,{t}^{10}+67\,{t}^{11} + \cdots
\end{eqnarray}
This series is a quite ``checkered'' one. Furthermore, its coefficients
depend very much on parameter $\, \epsilon\, $. In contrast with 
generating function (\ref{genptfixe}), the 
generating function $\, H^{\rm real}_{\epsilon} \, \, $ 
has {\em no universality property} in
$\epsilon$. This series does not take into account 
the topological invariance in complex projective space : it just tries to
describe the dynamical system in the real  space.
This series $\, H^{\rm real}_{\epsilon}\, $ corresponds to the ``complexity'' 
as seen on the phase portrait of Fig.~(\ref{f:fig3}).
One sees here the quite drastic opposition between the 
notions well-suited to describe transformations in
complex projective spaces and the ones aiming at describing
 transformations in real variables.

\subsection{Dynamical zeta functions for $\alpha=0$, $\epsilon$ non generic}

To further investigate the identification
of these two notions (Arnold 
complexity-topological entropy), we now perform similar 
calculations (of fixed points and associated
 zeta dynamical functions) for  $ \epsilon\, = \, 1/m\, $
with  $\,m \, \ge \, 4\, $ 
and  $ \epsilon\, = \, (m-1)/(m+3)\, $ with $\, m\, \ge 7 $ odd.

The calculations have been performed  for  
$ \, \epsilon\, = \, 1/m\, $
for $\, m \, = \, 4\, ,\,  5\, , \, 7 $ and $\, 9$,
giving the expansion of $\, H_{\epsilon}(t) \,$ up to order eleven.
For $m\, = \, 4\, $ this gives :
\begin{eqnarray}
\label{G4}
H_{1/4}(t) \, = \, \,  \, t+{t}^{2}+4\,{t}^{3}+5\,{t}^{4}+11\,{t}^{5}
+10\,{t}^{6}+22\,{t}^{7}+29\,{t}^{8}+49\,{t}^{9}
 +71\,{t}^{10}+111\,{t}^{11} + \cdots
\end{eqnarray}
for $\, m\, = \, 5\, $  :
\begin{eqnarray}
\label{G5}
H_{1/5}(t) \, = \, \, t+{t}^{2}+4\,{t}^{3}+5\,{t}^{4}+11\,{t}^{5}
+16\,{t}^{6}+22\,{t}^{7}+
37\,{t}^{8}+58\,{t}^{9}
 +91\,{t}^{10}+144\,{t}^{11} + \cdots
\end{eqnarray}
for  $m\, = \, 7\, $  :
\begin{eqnarray}
\label{G7}
H_{1/7}(t) \, =  \, \, \,
t+{t}^{2}+4\,{t}^{3}+5\,{t}^{4}+11\,{t}^{5}+16\,{t}^{6}
+29\,{t}^{7}+45\,{t}^{8}+67\,{t}^{9}
 +111\,{t}^{10}+177\,{t}^{11} + \cdots
\end{eqnarray}
and for  $m\, = \, 9\, $  :
\begin{eqnarray}
\label{G9}
H_{1/9}(t) \, =  \, \, \,
t+{t}^{2}+4\,{t}^{3}+5\,{t}^{4}+11\,{t}^{5}+16\,{t}^{6}
+29\,{t}^{7}+45\,{t}^{8}+76 \, t^9\, + \, 121 \, t^{10}\, 
+\, 188 \, t^{11} + \cdots
\end{eqnarray}

All these expressions are compatible with this single
expression of the $\zeta$ function :
\begin{eqnarray}
\label{zetam}
\zeta_{1/m}(t)  \, = \, \, {{1\, -t^2 } \over {1\, -t\, -t^2\, + t^{m+2}}} 
\end{eqnarray}
We conjecture that this expression is exact at every order
and for every value of $m  \ge 4$. Again all the singularities
of this expression coincide with those of generating
function corresponding to the Arnold complexity
(see  Eq.~(\ref{complexspec})).

As far as functional relations relating $\zeta(t)$ and 
 $\zeta(\pm 1/t)$ are concerned,
recalling (\ref{avatar}),
one immediately verifies that $\, \widehat{\zeta}(t)\, $
corresponding to (\ref{zetam})  verifies the
simple functional relation :
\begin{eqnarray}
\label{Waouh}
t^{m+1} \cdot \widehat{\zeta}_{1/m}(t)\,  \, \, = \,\,\, \, \,
\widehat{\zeta}_{1/m}(1/t)\, , \qquad \,\,\, \hbox{or:} \qquad \quad 
\zeta_{1/m}(1/t) \, = \,
\, {{t^{m+1} \cdot \zeta_{1/m}(t)  } \over {t^{m+1} \cdot 
\zeta_{1/m}(t) \, - \,
\zeta_{1/m}(t)\, + \, 1 }}
\end{eqnarray}
Actually  $\, \widehat{\zeta}_{1/m}(t)\, $ has a very simple
$n$-th root of unity  form :
\begin{eqnarray}
\label{simpleisntit}
\widehat{\zeta}_{1/m}(t)  \, \,= \,\,\,\,
 {\frac {1-{t}^{2}}{t \cdot \left (1-{t}^{m+1}\right )}}
\end{eqnarray}
Also note that when $m$ is odd, and only in that case,
 $\, \widehat{\zeta}_{1/m}(t)\, $
{\em also satisfies} the functional relation :
\begin{eqnarray}
\label{also}
t^{m+1} \cdot \widehat{\zeta}_{1/m}(t)  \,\, \, = 
\, \, \,  -\,\, \widehat{\zeta}_{1/m}(-1/t)
\end{eqnarray}
No simple functional relation, similar to (\ref{gg}), can 
be deduced on $H(t)$.

Similar calculations can also be performed for the second set of 
non-generic values of $\, \epsilon \, $, namely  
$\epsilon =\, (m-1)/(m+3)$ with $m \ge 7$, $m$ odd. 
For $m=7$, that is $\epsilon =\, 3/5$, one gets, up to order eleven, the 
{\em same expansion} as Eq.~(\ref{G7}) :
\begin{eqnarray}
\label{G7odd}
H_{3/5}(t) \, =  \, \,
t+{t}^{2}+4\,{t}^{3}+5\,{t}^{4}+11\,{t}^{5}+16\,{t}^{6}+29\,{t}^{7} 
+45\,{t}^{8}+67\,{t}^{9}
 +111\,{t}^{10} +177\,{t}^{11}+ \cdots
\end{eqnarray}
suggesting, again, the dynamical zeta function :
\begin{equation}
\label{zeta7new}
\zeta_{3/5}(t)  \, = \, {{1\, -t^2 } \over {1\, -t\, -t^2\, + t^{9}}} 
\end{equation}

For $m=9$, that is $\epsilon =\, 2/3$, one gets 
\begin{eqnarray}
\label{G9odd}
H_{2/3}(t) \, =  \, \,
t+{t}^{2}+4\,{t}^{3}+5\,{t}^{4}+11\,{t}^{5}+16\,{t}^{6}+29\,{t}^{7} 
+45\,{t}^{8}+76\,{t}^{9}
 +121\,{t}^{10} +177\,{t}^{11} \, + \cdots \nonumber
\end{eqnarray}
A compatible zeta function could be\footnote{The series is not large
enough to confirm this form. A first simple analyzis seems to show that the
next terms are $ \cdots + 296 \, t^{12} \, 
+ 469 \, t^{13}  \, + 785 \, t^{14} \, + \cdots    $.} :
\begin{eqnarray}
\label{zeta9new}
\zeta_{2/3}(t)  \, = \,\,  {{1\, -t^2 -t^{11}-t^{12}-t^{13}} \over {1\, -t\,
-t^2\, + t^{11}}}
\end{eqnarray}
This form is not the same as Eq.~(\ref{zetam}), however
it has the same poles. 

Comparing these rational expressions for the 
dynamical zeta function ((\ref{conjec}), (\ref{zetam}), ...),
and the  rational expressions for the generating functions of the
Arnold complexity ((\ref{betabeta1}), (\ref{betabeta2}),
(\ref{betabeta3}), ...)
for the generic, and non-generic, values of $\, \epsilon\, $,
one sees that
one actually has the same singularities
in these two sets of generating functions
(note that $t$ has to be replaced by $x^2$ since 
$\, k_{\epsilon}\, $ is associate to transformation $\, K^2$
and not $\, K$).
The identification between the Arnold complexity 
and the  (exponential of the) topological entropy 
is thus valid {\em for generic values of 
$\epsilon$, } and {\em even for non-generic ones}.

It is worth noticing that, due to the topological character
of the dynamical zeta function, these results are of course
not specific of the $y$ and $z$ representation of the
mapping (\ref{yz}) but are also valid for the $(u\, , v)$ 
representation (\ref{uv}) : in particular the exact expressions
of the dynamical zeta functions (namely (\ref{conjec}), (\ref{zetam})),
remain unchanged and, of course, the denominators of the complexity
generating functions are also the same for generic, or  non-generic,
values of $\, \epsilon$.

The {\em local area preserving} property  {\em in the neighborhood
of all the fixed points of} $k_{\epsilon}^n$ previously
noticed for $\alpha=0$, $\epsilon$ generic, is also verified 
for these non generic values of  $\epsilon$.

\subsection{Dynamical zeta functions for $\alpha \ne 0$}

This (generic) identification is not 
restricted to $\, \alpha \, = \, 0\, $.
One can also consider mapping (\ref{uv}) for arbitrary 
values of $\, \alpha \,$ and $\, \epsilon \, $ 
and calculate the successive fixed points.
Of course, as a consequence of the higher complexity
of the  $\, \alpha \, \ne \, 0\, $ situation 
(as shown in  section~\ref{secB}, the complexity immediately jumps
from $\, 1.61803 \cdots \, $ to $\, 2.14789 \cdots \, $)
the number of successive fixed points is drastically increased and the
calculations cannot be performed up to order eleven anymore. 
In the generic  case, the expansion 
of the generating function $H(t)$ of the number of fixed points 
can be obtained up to order seven :
\begin{eqnarray}
H_\epsilon^\alpha \,=  \,  \,\,
2\,t+2\,{t}^{2}+11\,{t}^{3}+18\,{t}^{4}+47\,{t}^{5}+95\,{t}^{6}\,  \, + \,
 212\,{t}^{7} \,+ \, 
\cdots 
\end{eqnarray}
One has two fixed points for $k$, no new fixed points for $k^2$,
three sets of three new fixed points for $k^3$ (giving $ 3 \times 3 \,
+ \, 2\, = \, 11$ fixed points),  four 
sets of four new fixed points for $k^4$ (giving $ 4 \times 4 \,
+ \, 2\, = \, 18$ fixed points), nine sets of five new fixed points 
for $k^5$ (giving $ 9 \times 5 \,
+ \, 2\, = \, 47$ fixed points), fourteen sets of six new fixed points 
for $\, k^6$ (giving $ \, 14 \times 5 \,+ \, 3 \times 3 \,
+ \, 2\, = \, 95$ fixed points).
This expansion corresponds to 
the following order seven expansion
for the dynamical zeta function :
\begin{eqnarray}
\zeta_\epsilon^\alpha(t)  \, = \,\, 1\, +2\,t\, +3\,{t}^{2}\, +7\,{t}^{3}\,
+15\,{t}^{4}+32\,{t}^{5}\, 
+69\,{t}^{6}\,  + \,
 148 \,{t}^{7}\,+ \,  \cdots 
\end{eqnarray}
thus yielding to the following rational expression
for the dynamical zeta function :
\begin{eqnarray}
\label{zetaalpha}
\zeta_\epsilon^\alpha(t)  \, = \,\, 
{\frac {(1-t^2) \cdot (1+t)}{1-t-2\,t^2\, -\,t^3}} \, = \,\,
{{(1\, -\, x^2) \cdot (1\, +\, x^2)^2} \over 
{(1\, -\, x\, -\, x^3) \cdot (1\, +\, x\, +\, x^3)}} \, \qquad \, \, 
\hbox{with :} \qquad \, \,   t \, = \, x^2
\end{eqnarray}
This expression can also be written :
\begin{eqnarray}
\zeta_\epsilon^\alpha(t)  \,  \,\, = \,\,  \,\, 
{\frac {\left (1\, -\, t^2\right ) \cdot 
\left (1+t\right )}{1\, -t \cdot \left (1+t\right )^{2}}}
\end{eqnarray}

Let us recall  the ``alternative'' zeta
function (\ref{avatar}). It verifies 
the simple functional relation :
\begin{eqnarray}
t^2 \cdot \widehat{\zeta}_\epsilon^\alpha(t) 
\cdot \widehat{\zeta}_\epsilon^\alpha(-t)\,\,\, = \,\, \, \, 
-\,  \widehat{\zeta}_\epsilon^\alpha(-1/t) 
\cdot \widehat{\zeta}_\epsilon^\alpha(1/t)\,
\end{eqnarray}
This new rational conjecture (\ref{zetaalpha})
corresponds to the following expression for  $H(t)$ : 
\begin{eqnarray}
H_\epsilon^\alpha(t) \,\, \,  = \,\, \,\,
{\frac {t \cdot  ( 2 \, + \, 3\, t^2 \, +  \, t^3 )}{\left
(1-{t}^{2}\right )
\cdot \left (1-t \, -2\,{t}^{2}\, -{t}^{3}\right )}}\, \,\,  \, \,
\end{eqnarray}

Comparing the denominators of Eq.~(\ref{zetaalpha}) and Eq.~(\ref{alpeps}),
one sees that, like for the case $\alpha=0$, 
there is an  identification between 
the Arnold complexity and the  (exponential of the) topological entropy
\begin{equation}
\lambda \, =\, \,  h
\end{equation}
Heuristically, this  identification can be understood as follows. 
The components of $k^N$, namely $y_N$ and $z_N$, are of the form
$\, P_N(y,z)/Q_N(y,z)\, $ and $\, R_N(y,z)/S_N(y,z)\, $,
where $\, P_N(y,z)\, $,  $\, Q_N(y,z)\, $,  $\, R_N(y,z)\, $
and  $\, S_N(y,z)\, $ are polynomials of degree asymptotically growing
like $\, \lambda^N$. The Arnold complexity amounts to taking the
intersection of the $N$-th iterate of a line (for instance
a simple line like  $\, y=y_0$
where $\, y_0$ is a constant)
with another simple (fixed) line (for instance
  $\, y=y_0$ itself
or any other simple line or any {\em fixed} algebraic curve).
For instance,  let us consider the $N$-th iterate of the $\, y=y_0$
line, which can be parameterized as :
\begin{eqnarray}
y_N \, = \, {{ P_N(y_0,z)} \over {Q_N(y_0,z)}}\, , \qquad \quad
z_N \, = \, {{ S_N(y_0,z)} \over {T_N(y_0,z)}}\, ,
\end{eqnarray}
with  line   $\, y=y_0$ itself.
The number of intersections, which are 
the solutions of $\, \, P_N(y_0,z)/Q_N(y_0,z) = y_0$,
grows like the degree of  
$\, \, P_N(y_0,z)\, - \, Q_N(y_0,z) \cdot y_0\, $:
 asymptotically it grows like $\, \simeq \, \lambda^N$. On the 
other hand the calculation of the topological entropy
corresponds to the number of fixed points of $k^N$,
that is to the number of intersection of the two curves:
\begin{eqnarray}
 P_N(y,z)\, - \, Q_N(y,z) \cdot y\,= \, \, 0\, , \qquad \quad
R_N(y,z)\, - \, S_N(y,z) \cdot z\,= \, \, 0\, 
\end{eqnarray}
which are two curves of degree growing  
asymptotically  like $\, \simeq \, \lambda^N$.
The number of fixed points is obviously bounded by 
 $\, \simeq \, \lambda^{2\, N}\, $ but one can figure out that it should
(generically) grow like  $\, \simeq \, \lambda^N$. This is fully
confirmed by our exact calculations.

The eulerian product Weyl-decomposition (\ref{Weyl}) 
of the dynamical zeta function
(\ref{zetaalpha}) corresponds to the following numbers
of $r$-cycles : 
$N_1\, = \, 2\, , \, N_2\, = \, 0\,, \,  N_3\, = 3\, , $
$\,  N_4\, = 4\, ,\,  N_5\, = 9\,,$$  \,\,  N_6\, = 14\, , 
\,\,  N_7\, = 30\, , \,  N_8\, = 54\, ,$$ \, \, N_9\, = 107\, , 
\,  N_{10}\, = 204\, , $$\, \, N_{11}\, = 408\, , 
\,  N_{12}\, = 25\,,$$\, \, N_{13}\, = \,1593\, , 
\, \, N_{14}\, = \,3162$.

\subsection{Dynamical zeta functions for $\alpha \ne 0$ with $\epsilon$
non-generic}

For a ``non-generic'' value of $\epsilon$ when  $\alpha \ne 0$,
namely $\epsilon = 1/2$, the expansion
of the generating function $H(t)$ and of the  dynamical zeta function
read respectively :
\begin{eqnarray}
&&H_{1/2}^\alpha(t) \, = \, \, 2\,t+2\,{t}^{2}+11\,{t}^{3}+18\,{t}^{4}
+47\,{t}^{5}+95\,{t}^{6}+198\,t^7\, + \cdots   \nonumber \\
&&\zeta_{1/2}^\alpha (t)  \, = \,\, 1+2\,t+3\,{t}^{2}+7\,{t}^{3}+15\,{t}^{4}
+32\,{t}^{5}+69\,{t}^{6}+146\,{t
}^{7}\, + \cdots   \nonumber 
\end{eqnarray}
A possible rational expression
for the dynamical zeta function is
for instance :
\begin{eqnarray}
\label{zetaalpha1sur2}
\zeta_{1/2}^\alpha \,= \,
{\frac
{1+t-{t}^{7}}{1-t-{t}^{2}-2\,{t}^{3}-{t}^{4}-2\,t^5\, -t^6-t^7}} \,
\,= \, \, \, \,  
{\frac {1\, +t \cdot \left (1\, -{t}^{6}\right )}{1\, -\, 
t \cdot (1\,-t\, +t^2 )
\cdot (1\, +t\, +t^2 )^2}} \, 
\end{eqnarray}
This last  result has to be compared with (\ref{alpeps1sur2}).

For another ``non-generic'' value of $\epsilon$ when  $\alpha \ne 0$,
namely $\epsilon = 1/3$ the expansion
of the generating function $H(t)$ and of the  dynamical zeta function
read respectively :
\begin{eqnarray}
&&H_{1/3}^\alpha(t) \, \, = \, \, \, \,2\,t+2\,{t}^{2}+11\,{t}^{3}+18\,{t}^{4}
+42\,{t}^{5}+83\,{t}^{6}+177\,{t}^{7}+ \cdots   \nonumber \\
&&\zeta_{1/3}^\alpha(t)  \, \, = \,\, 
 1+2\,t+3\,{t}^{2}+7\,{t}^{3}+15\,{t}^{4}+31\,{t}^{5}+65\,{t}^{6}
+136\,{t}^{7} + \cdots  \nonumber 
\end{eqnarray}
A possible rational expression
for the dynamical zeta function is
for instance : 
\begin{eqnarray}
\label{zetaalpha1sur3}
\zeta_{1/3}^\alpha(t) \, = \, \, \, {\frac
{1+t}{1-t-{t}^{2}-2\,{t}^{3}-{t}^{4}-{t}^{5}}} \, = \, \,
 \,  \, \,{\frac {1+t}{1\, -\, t \cdot  (1\, +\, t^2 )
\cdot  (1\, +t\, +t^2 )}}
\end{eqnarray}

This last  result with 
has to be compared with (\ref{alpeps1sur3}).
These results\footnote{However for the non-generic value
of $\, \epsilon \,$, $ \, \epsilon\, = \, 3/5$, we do not have enough
coefficients in the expansion of
the dynamical zeta function to actually compare
 it with (\ref{g35}).} are again in agreement with an 
Arnold-complexity-topological-entropy identification.

The {\em local area preserving} property   in the neighborhood
of all the fixed points of $k_{\alpha, \epsilon}^n$ previously
noticed for $\alpha=0$, is also verified 
for $\alpha \ne 0$  for (\ref{uv}) {\em for  generic values of}
 $\epsilon$ {\em generic as well as these
 non generic values of}  $\epsilon$.


\section{Comments and speculations}

Based on analytical and semi-numerical calculations we have 
conjectured rational expressions with integer coefficients
for the generating functions
of the complexity and for the dynamical zeta functions
for various values of the parameters of a family 
of birational transformations. According to these conjectures,
the Arnold complexity and the  exponential of the
topological entropy are
algerbraic numbers. Moreover, these two numbers are {\em equal}
for all the values of the parameters.

From a general point of view,
rational zeta dynamical functions  (see for
instance~\cite{Gu70,Ba91,Ru91}) are known in the literature
 through theorems where the dynamical
systems are asked to be {\em hyperbolic}, 
or through combinatorial
proofs using symbolic dynamics arising from Markov
partition~\cite{Manning}
 and even, far beyond these
frameworks~\cite{Fried87}, for the so-called ``isolated 
expansive sets''(see~\cite{Fried87,Conley} for a definition of the 
isolated  expansive sets).
There also exists an explicit  example of a rational zeta
dynamical function but only in the case 
of an {\em explicit linear} dynamics
on the torus $\, R^2/Z^2\, $,  deduced from 
an $SL(2,Z)$ matrix, namely the cat map~\cite{ASY,AA67} 
(diffeomorphisms of the torus) :
\begin{eqnarray}
 A \, = \, 
\left [\begin {array}{cc}
 2&1\\
\noalign{\medskip}1&1
\end {array}
\right ] \, , 
\, \,\qquad  B \, =
 \,\left [\begin {array}{cc} 1&0\\
\noalign{\medskip}0&1
\end {array}
\right ]
\, , \qquad \zeta \, = \, \,
 {{{\rm det}(1-z \cdot B)} \over {{\rm det}(1-z
 \cdot A)}} \, = \, \, {{(1-z)^2} \over { 1-3 \cdot z +z^2}}
\end{eqnarray}
Note that   golden number singularities
for complexity growth generating 
functions have already been encountered (see equation (7.28) in
~\cite{BoMa95} or equation (5) in ~\cite{HiVi97}).
In our examples, we are not in the context where the known general
theorems can apply straightforwardly. The question of the
demonstration of the rationality of zeta functions we have 
conjectured remains open.

In the framework of a ``diffeomorphisms of the torus'' interpretation,
the degree of the denominator of a  rational
 dynamical zeta function 
gives a lower bound of the dimension $\, g$ 
of this ``hidden'' torus $ C^g/Z^g$.
On expression (\ref{zetam}) valid for $\, \alpha=0\, $ and
 $\epsilon \, = \, 1/m$, one notes that  dimension $g$
grows linearly with $m$. The iteration of 
some birational transformations which
densify Abelian surfaces (resp. varieties) has been seen
to correspond to  
polynomial growth of the calculations~\cite{BoMaRo95}.
Introducing well-suited variables $\theta_i$ ($i=1, \, \cdots g$)
to uniformize
the Abelian  varieties the iteration of these birational  transformations
 just corresponds to a shift\footnote{
This ``diffeomorphisms of the torus'' interpretation
is quite obvious on figure 2 of~\cite{BoMa95}.} $\,\theta_i  
\, \, \rightarrow \, \, \theta_i\, + \,
n \cdot \eta_i$. For such polynomial growth
situations, matrix $A$ can be thought as the
Jordan matrix associated with this translation, its
characteristic polynomial yielding eigenvalues equal to $\, 1$.

Many denominators
of  rational zeta functions encountered here
are of the form : $\, 1\, -\, t \cdot Y(t)\, $ where $\, Y(t)\, $ is product
 of cyclotomic polynomials~\cite{cyclo,cyclo1}. We have encountered :
\begin{eqnarray}
&&Y(t) \, = \, \, \, (1+t)\quad  (\hbox{resp.} \quad (1+t)^2)   \,
 \, \qquad \hbox{for} \qquad \alpha \, =
\, 0 \quad  (\hbox{resp.} \quad \alpha \,
\ne  \, 0) 
\qquad \hbox{and} \, \quad \, \epsilon  \, \,\hbox{generic}\, ,  \nonumber \\
&&Y(t) \, = \, \, \, {{(1+t^3)} \over {1+t}}   
\cdot {{(1-t^3)^2} \over {(1-t)^2}} \, \quad 
 (\hbox{resp.} \quad  (1+t^2) \cdot {{(1-t^3)} \over {1-t}} )
 \qquad \hbox{for} \qquad  \alpha \,
\ne  \, 0 \qquad \hbox{and } \, \, \epsilon  \,= \, {{1} \over {2}}\, \quad 
 (\hbox{resp.}  \epsilon  \,= \, {{1} \over {3}}) \nonumber 
\end{eqnarray}
More generally
the  rational dynamical zeta functions, or the  rational functions
$G(q,\, x)\, $
encountered here,
are of the form : $(1+X(z))/(1-Y(z))$ (for $\, G\, $)
or $(1-X(z))/(1-Y(z))$ (for $\zeta$) where $X(z)$ and $Y(z)$ have
 some kind of decomposition on {\em cyclotomic
 polynomials} :
\begin{eqnarray}
X(z) \, = \, \, \sum_r  z^r \cdot \Pi_m P^{(r)}_m(z)
\quad  \quad \hbox{with} \quad  \quad P^{(r)}_m(z) \quad \quad
\hbox{cyclotomic}\quad 
\hbox{polynomials}
\end{eqnarray}
This is particularly obvious on expressions (\ref{betabeta3})
but also on expressions  (\ref{betabeta2}), or (\ref{zetaalpha1sur2}),
or even (\ref{zetam}). We do not have yet any $l$-adic cohomology
interpretation (see for
 instance~\cite{W} page 453) of this cyclotomic
 polynomials ``encoding'' of the zeta functions or  the complexity
functions $\, G(q, \, x)\,$.
Most of these rational expressions for zeta functions
satisfy very
simple functional 
relations 
but one also expects, for (\ref{zetaalpha1sur2}) or
(\ref{zetaalpha1sur3}) for instance, more involved
but, still simple, functional 
relations similar to the ones obtained by Voros in~\cite{Vo81}.
Many of the generating functions $\, G(q, \, x)\,\, $ can also 
be seen to satisfy 
simple functional relations relating $G(q,\, x)$ and $\, G(q, \, 1/x)$. 
This will be detailed elsewhere\footnote{For instance the generating function
of the degrees $\, g(x)$ given by equation (5) in~\cite{HiVi97} verifies
$\,\, g(x)\, + \, g(1/x) \, =\, \, 1$.}.

In practice
 it is numerically easier to get the generating functions of Arnold 
complexity than getting the dynamical zeta functions.
If one assumes the rationality
of the dynamical zeta function and the identification
 between Arnold complexity and (exponential of the)
topological entropy, getting the generating functions of Arnold 
complexity  is a simpler way to ``guess'' the denominator 
of the dynamical zeta functions.

The analysis developed here can be applied to a very large set of birational 
 transformations of an {\em arbitrary} number of
 variables, always leading {\em rational} generating 
functions~\cite{BoMa95,AbAngBoHaMa98}.
Moreover, these generating 
functions are always simple rational expressions with 
integer coefficients (thus yielding {\em algebraic numbers} for the Arnold
 complexity). They even have the previously mentioned 
``cyclotomic encoding''.
At this point, the question can be 
raised\footnote{After~\cite{BoMa95}.} to see if the iteration of
 {\em any} birational transformation of an arbitrary number of
 variables always yields rational generating functions 
for the Arnold complexity. We have even found rational
 generating functions of Arnold 
complexity for rational transformations {\em which are not 
birational} (see (7.7) and (7.28) in~\cite{BoMa95}) : any 
proof of these rationalities should not depend too heavily on
a naive {\em reversibility}  of the mapping~\cite{QuRo88}.

We have also calculated Lyapunov exponents~\cite{AbAngMa98}
in order to study the {\em metric entropy}.
These numerical calculations will be detailed
 elsewhere~\cite{AbAngMa98}
for transformation (\ref{yz}) for $\epsilon= .52$. These results share quite
small values of the  Lyapunov exponents, the largest of which being
much smaller than the topological entropy.
We thus infer that, in this very example, the metric entropy is
 much smaller than the topological entropy. We have here an opposition
between topological concepts originating from 
complex projective spaces and the metric concepts of real analysis.
The ``non-topological'' complexity measures do not seem to be able
to identify
with the previous topological and algebraic quantities.
On the birational examples studied here, the metric entropy
does not seem to share the same algebraic values
as the topological complexity measures.

\acknowledgments{One of us (JMM) would like to thank 
P. Lochak and J-P. Marco for illuminating discussions
on dynamical systems. We thank M. Bellon and 
C. Viallet for complexity discussions.
We thank B. Grammaticos and 
A. Ramani for many discussions on the non generic values
of $\epsilon$.
S. Boukraa would like to thank 
the CMEP for financial support.
}

\appendix

\section{Correspondence between transformation $K$
acting on $\, q \times q$ matrices and the $(u\, , \, v)$
transformation $K_{\alpha,\epsilon}$}
\label{AppendixA}

In  previous papers~\cite{BoMaRo94,BoMaRo93c}, it has been noticed that the
successive {\em even} iterates of $\, K\, $, acting on an initial
matrix $M$, actually belong to 
a plane which contains 
matrices $M$ , $\, K^2(M)\, $ and  $K^4(M)$, or equivalently 
a fixed matrix $P$ (see~\cite{BoMaRo94}) :
\begin{eqnarray}
K^{2\, n} (M) \, = \, \, c_0 \cdot M \, + \,c_1 \cdot K^2(M) \,
+\, c_2 \cdot P \, \nonumber
\end{eqnarray}
for any integer $n$ (even relative integer).
In fact one even has the following property : any point of
the plane containing $M$, $\, K^2(M)$ and $P$ is transformed, by
the even iterates 
$K^{2\, n}$, into another point of this plane.
This can be used to define the two dimensional mapping 
$k_{\epsilon,\alpha}$ compatible with mapping $\, K$.
Let 's $M$ be an arbitrary $q \times q$ matrix, and let us 
define\footnote{Note a
miss-print in~\cite{BoMaRo94} : in equation (6.35) $u_n/v_n$ should be
replaced by  $v_n/u_n$ yielding (see (\ref{aaM}) and (\ref{aM})) :
$v_n/u_n\, = \, \, $ $ x_3 \cdot  \cdot x_5 \cdot x_7 \cdots x_{2n-1}
\cdot v_1/u_1 \, $.} :
\begin{eqnarray}
&&u_n(M)   \,   =  \, x_2 \cdot x_4 \cdot x_6 \cdots x_{2n-2} \cdot
u_1(M) \,,  \qquad 
v_n(M)   \,   =  \, x_2 \cdot x_3 \cdot x_4 \cdot x_5 
\cdot x_6 \cdots x_{2n-1} \cdot
v_1(M)  \\
\label{aaM}
&&\hbox{where :}\qquad u_1(M)     \,   =  \, \frac {x_0}{\rho_2} \,,  \qquad 
v_1(M)     \,   = \, \, - \frac {x_0 x_1}{\lambda_2} \, , \qquad
\hbox{and:}
\qquad
\alpha(M)  \,   =  \, \, - \frac{ \rho_1  \rho_2}{\lambda_2} \, , \qquad 
\epsilon(M)  \, =\,  \,  \frac{\lambda_1 - \lambda_2}{\lambda_2} 
\label{aM}
\end{eqnarray}
where $\,\,\, x_n \, =\, \,  {\rm det}(\widehat{K}^m(M)) \cdot {\rm
det}(\widehat{K}^{n+1}(M)) \, \, \, $ with $\, \, \widehat{K} \,  \, = t \cdot
\widehat{I}$,
and  $ \,\widehat{I}(M)\, = \, M^{-1} \,$
and\footnote{Note that the $\lambda_i$'s and $\rho_i$'s are not
exactly
the same as the ones given in~\cite{BoMaRo94}, in equations (6.13) and
(6.14) : the  $\lambda_i$'s and $\rho_i$'s in~\cite{BoMaRo94} are
homogeneous expressions and the  $\lambda_i$'s and $\rho_i$'s we introduce here
are inhomogeneneous true invariants which can be deduced from the
ones in~\cite{BoMaRo94} dividing them be $q_0$ or $q_1$.} :
\begin{eqnarray}
\lambda_1 &=& \, \,\, \frac {x_0 x_2 \cdot (x_1 x_3 -1)}{x_2-1}\,,
\qquad 
\lambda_2 =\, \,\, \frac {x_1 \cdot (x_0 x_2 -1)}{x_1-1} \, , \nonumber \\
\rho_1 &=&\, \,\, \frac {x_1 x_2 x_3 + x_1 x_2 - x_1 -1}{x_2 -1}\, , \qquad 
\rho_2 =\, \,\, \frac {x_0 x_1 x_2 + x_0 x_1 - x_0 -1}{x_1 -1}\nonumber 
\end{eqnarray}
Then one has the $K^2$ invariance of $\alpha$ and $\epsilon$ :
\begin{eqnarray}
\alpha(M)    = \, \alpha(K^2(M)) \,,  \qquad 
\epsilon(M)  =\,  \epsilon(K^2(M))
\end{eqnarray}
and $k_{\alpha,\epsilon}\, $ can be seen as a representation of 
$\, K^2$  :
\begin{equation}
\nonumber
\left( u(M),v(M) \right)\,\,   = \, \, \, k_{\alpha,\epsilon} 
	\left( u(K^2(M)),v(K^2(M) \right)
\end{equation}
where $\, \alpha\, $ and $\, \epsilon \, $ are precisely the values
given by Eq.~(\ref{aM}). Transformation $\, k_{\alpha,\epsilon} \, $
reads :
\begin{eqnarray}
k_{\alpha,\epsilon}: \qquad (u\, , \, v) \quad \longrightarrow \qquad 
(U\, , \, V) \, = \, \, ( {{ v+u-u\cdot v} \over {v}}\, , \,\,  \, 
 U \cdot \alpha \, 
+ \,1 \, + \, \epsilon\,  -\, {{ u\, +v\, -u\cdot v} \over {u}})
\end{eqnarray}

In the $\alpha \, = \, 0\, $ case, this transformation simplifies
and one can introduce new variables $y$ and $z$ given by :
\begin{eqnarray}
y \, = \, \, \, v-1 , \qquad z \, = \, \, {{ (1-u) \cdot (1-v) \cdot
v} \over {u \cdot (v-1)}}
\end{eqnarray}
With these new variables, $k_{\alpha,\epsilon}$ reads :
\begin{eqnarray}
k_{\epsilon}: \qquad (y\, , \, z) \qquad \longrightarrow \qquad 
(z +1 - \epsilon\, , \, \,
 y \cdot
 \frac{z-\epsilon}{z + 1} )
\end{eqnarray}
For $\alpha \, = \, \epsilon \, = \, 0  $, transformation 
$\, k_{\alpha,\epsilon}$ 
is integrable~\cite{BoMaRo94}
the invariant being (see (6.38) in~\cite{BoMaRo94}) :
\begin{eqnarray}
{\cal I} \, = \,  {{ (1-u) \cdot (1-v) \cdot
v} \over {u }}
\end{eqnarray}
This algebraic expression is of course only 
well-suited for  $\epsilon = 0$. The variable $z$ amounts to 
considering ${\cal I}/(v-1)$ for arbitrary $\epsilon$'s.

\section{Factorization scheme for (\ref{K})}
\label{AppendixB}

For $q \times q$ matrices ($q \ge 3$)
 the factorizations corresponding to the iterations of $K$  read:
\begin{eqnarray}
\label{rappel}
&&f_1\, = \, det(M_0) \, , \quad 
M_1\,= \, K(M_0) \,, \quad
f_2\  = \, \frac{det(M_1)}{f_1^{q-2}} \, , \quad
M_2\,= \, \frac{ K(M_1) }{ f_1^{q-3}} \, , \quad
f_3\, = \, \frac{det(M_2)}{f_1 \cdot f_2^{q-3} } \, ,\quad 
M_3\,= \, \frac{K(M_2)}{f_2^{q-3}} \,, \nonumber \\
&&f_4\, = \, \frac{det(M_3)}{f_1^{q-1} \cdot f_2 \cdot f_3^{q-2}} \, , \qquad
M_4\,= \, \frac{K(M_3)}{f_1^{q-2} \cdot  f_3^{q-3}} \,, \qquad 
f_5\, = \, \frac{ det(M_4)}{f_1^2 \cdot f_2^{q-1} \cdot f_3 
                \cdot f_4^{q-2} } \, , \qquad 
M_5\,= \, \frac{K(M_4)}{ f_1 \cdot  f_2^{q-2}\cdot
    f_4^{q-3}} \, , \nonumber \\
&& f_6\, = \, \frac{det(M_5)}{f_1^{q-2} \cdot 
f_2^2 \cdot f_3^{q-1} \cdot f_4 \cdot 
        f_5^{q-2} } \, ,\qquad 
M_6 \, = \, \frac{K(M_5)}{f_1^{q-3} \cdot  f_2 \cdot  
    f_3^{q-2}\cdot  f_5^{q-3}} \, , \, \qquad 
f_7\, = \, \frac{ det(M_6)}{f_1 \cdot f_2^{q-2} \cdot 
        f_3^2 \cdot f_4^{q-1} \cdot f_5 \cdot f_6^{q-2} } \, , \nonumber \\
&&  M_7 \,= \, \frac{K(M_6)}{f_2^{q-3} \cdot  f_3 \cdot  
        f_4^{q-2}\cdot  f_6^{q-3}} \quad \quad \cdots  
\end{eqnarray}
and for arbitrary $n$ :
\begin{eqnarray}
\label{detIV}
det(M_n)\, \, &=&\, \,\, \,  f_{n+1} \cdot (f_{n}^{q-2} \cdot f_{n-1} 
\cdot f_{n-2}^{q-1} \cdot f_{n-3}^2 )\cdot (f_{n-4}^{q-2}
 \cdot f_{n-5} \cdot f_{n-6}^{q-1} 
\cdot f_{n-7}^2) \cdots f_{1}^{\delta_n} \, ,  \\
\label{KIV}
K(M_n)\, \, &=&\, \,\, \, M_{n+1} \cdot ( f_{n}^{q-3} \cdot f_{n-2}^{q-2} \cdot
 f_{n-3} )
 \cdot (f_{n-4}^{q-3} \cdot f_{n-6}^{q-2} \cdot f_{n-7} ) \cdots f_{1}^{\mu_n}
\end{eqnarray}
where $\mu_n=q-3$ for $n=1$ (mod 4), $\mu_n=0$ for $n=2$ (mod 4),
 $\mu_n=q-2$ for $n=3$ (mod 4) and $\mu_n=1$ for $n=0$ (mod 4)
and $\delta_n$ also depends on the truncation.
Factorization relations
{\em independent of $q$}, {\em occur} :
\begin{eqnarray}
\label{KhatIValphq=3}
 {{K(M_n)} \over {det(M_n)}} \, = \, \, 
 \,\,\,\, \, \,
{{ M_{n+1} \over {
f_{n+1} \cdot f_{n} \cdot f_{n-1}
 \cdot f_{n-2} \cdot f_{n-3} \cdot f_{n-4}  \cdots }}}
\end{eqnarray}

Let us introduce~\cite{BoMaRo94,BoMaRo93c}
 the generating functions $\alpha(x)$ and $\beta(x)$
 of the degree of the
$det(M_n)$'s and $f_n$'s.
Their exact expressions read:
\begin{eqnarray}
\label{albexxIV}
\alpha(x) \, = \, \,\, {\frac {q}{1+\, x}}
+ \, {\frac {q^2 \cdot  x \cdot 
\left (1 + \, x^{2} \right )}{(1-x)(1+x)(1-x-x^3)}}, \qquad \qquad 
\beta(x) \, = \,\,\, \,\, {\frac {q \cdot x \cdot 
\left (1+x^{2}\right )}{1-x-x^3}} 
\end{eqnarray}
It is clear that one has an exponential growth
of exponents $\alpha_n$'s, $\beta_n$'s, $\mu_n$'s and $\nu_n$'s: these
coefficients grow like 
$\, \lambda^n$ where $\, \lambda \sim 1.465 \cdots$

This displays the ``generic'' factorization scheme.
However, on various subvarieties like codimension one subvariety 
$\, \alpha \, = \, 0\, $, the  factorization scheme
can be modified as a consequence of additional factorizations
occurring at each iteration step,
 thus yielding a smaller value for the complexity
$\lambda \, $.

\subsection{Factorization scheme for $\alpha=0$, $\epsilon$ generic}
For $\alpha \, = \, 0\, $ the previous factorization
scheme becomes for $\, 3 \times 3 $ matrices\footnote{These results
can straightforwardly be generalized to $q \times q$ matrices, they
are just a bit more involved.} :
\begin{eqnarray}
\label{newnew}
&&f_1\, = \, det(M_0) \, , \quad 
M_1\,= \, K(M_0) \,, \quad
f_2\  = \, \frac{det(M_1)}{f_1} \, , \quad
M_2\,= \, \, K(M_1)  \, , \quad
f_3\, = \, \frac{det(M_2)}{f_1^2 \cdot f_2 } \, ,\quad 
M_3\,= \, \frac{K(M_2)}{f_1} \,, \nonumber \\
&&f_4\, = \, \frac{det(M_3)}{f_1 \cdot f_2 \cdot f_3} \, , \qquad
M_4\,= \, K(M_3) \,, \qquad 
f_5\, = \, \frac{ det(M_4)}{f_1^2 \cdot f_2^2 \cdot f_3^2 
                \cdot f_4 } \, , \qquad 
M_5\,= \, \frac{K(M_4)}{ f_1 \cdot  f_2 \cdot  f_3
    } \, , \nonumber \\
&& f_6\, = \, \frac{det(M_5)}{f_1 \cdot f_2^2 \cdot f_3 \cdot f_4 \cdot 
        f_5 } \, ,\qquad 
M_6 \, = \, \frac{K(M_5)}{  f_2 } \, , \, \qquad 
f_7\, = \, \frac{ det(M_6)}{f_1^2 \cdot f_2 \cdot 
        f_3^2 \cdot f_4^2 \cdot f_5^2 \cdot f_6 } \, , \qquad
 M_7 \,= \, \frac{K(M_6)}{f_1 \cdot  f_3 \cdot  
        f_4\cdot  f_5}  \nonumber \\
&&  f_8\, = \, \frac{det(M_7)}{f_1 \cdot f_2 \cdot f_3 \cdot f_4^2 \cdot 
        f_5  \cdot f_6 \cdot  f_7 } \, ,\qquad  \,  
M_8 \, = \, \frac{K(M_7)}{  f_4 } \, , \, \qquad 
f_9\, = \, \frac{ det(M_8)}{f_1^2 \cdot f_2^2 \cdot 
        f_3^2 \cdot f_4 \cdot f_5^2 \cdot f_6^2 \cdot f_7^2  \cdot f_8  } \, ,
\nonumber \\
&& M_9 \,= \, \frac{K(M_8)}{f_1 \cdot    f_2 \cdot f_3 \cdot  
        f_5 \cdot  f_6 \cdot  f_7 }\, , \quad \quad 
 f_{10}\, = \, \frac{det(M_9)}{f_1 \cdot f_2^2 \cdot f_3 \cdot f_4 \cdot 
        f_5  \cdot f_6^2 \cdot  f_7 \cdot  f_8 \cdot  f_9  } \,
, \, \, \cdots   \,  
\end{eqnarray}
and for arbitrary $n$ :
\begin{eqnarray}
\label{detIValphq=3even}
det(M_n)\,&=& \,\,\,  f_{n+1} \cdot (f_{n} \cdot f_{n-1}^2
 \cdot f_{n-2}^2 \cdot f_{n-3}^2 )\cdot (f_{n-4}
 \cdot f_{n-5}^2 \cdot f_{n-6}^2 
\cdot f_{n-7}^2) \,\, \cdots  \\
\label{KIValphq=3even}
K(M_n)\, &=&\,\,\, M_{n+1} \cdot ( f_{n-1} \cdot f_{n-2}
\cdot f_{n-3} ) \cdot (f_{n-5} \cdot f_{n-6}
\cdot f_{n-7} ) \,\, \cdots 
\end{eqnarray}
for $n$ even and :
\begin{eqnarray}
\label{detIValphq=3odd}
det(M_n)\,&=&\,\,\, f_{n+1} \cdot (f_{n} \cdot f_{n-1}
 \cdot f_{n-2} \cdot f_{n-3}^2 )\cdot (f_{n-4}
 \cdot f_{n-5} \cdot f_{n-6} 
\cdot f_{n-7}^2)\,\, \cdots \\
\label{KIValphq=3odd}
K(M_n)\, &=&\,\,\, M_{n+1} \cdot  f_{n-3} \cdot f_{n-7}
\cdot f_{n-11}  \cdot f_{n-15} \cdot f_{n-19}  \, \cdots 
\end{eqnarray}
for $n$ odd.

The exact expressions of the generating functions
 $\alpha(x)$ and $\beta(x)$ read\footnote{Result (\ref{albetqegal3})
 corresponds to a very simple expression for another generating
 function introduced in~\cite{BoMa95}, namely the function $\rho(x)$ (see
 for instance equation (8.12) in~\cite{BoMa95}).} :
\begin{eqnarray}
\label{albetqegal3}
\alpha(x) \, = \, \,\, {{3 } \over {1+x}} \, + \, {{ 3 \cdot \beta(x)}
\over {1-x^2}}
\, , \, \qquad \hbox{where :}\qquad
\beta(x) \, = \, \, 3 \cdot {\frac {x \cdot
 \left (1+x+{x}^{3}\right )}{1\, -\, x^2\, - \, x^{4}}}\, = \, \,
- \, 3\, + 3 \cdot (1+x)/(1-x^2-x^4) \, \,
\end{eqnarray}
It is important to note that factorization scheme
 (\ref{newnew}) is {\em actually stable, but of a slightly more general form,}
as compared to (\ref{rappel}), or 
the ones described in~\cite{BoMa95} : recalling the generating functions
$\eta(x)$ and $\phi(x)$ of the exponents that occur in 
the factorization scheme (see equation (8.6) and (8.10)  in~\cite{BoMa95}),
one must now introduce {\em two sets} of such  exponents generating functions,
$\eta_1$, $\phi_1$, $\eta_2$, $\phi_2$, 
in order to keep track of the parity of $n$,
and even split these four functions into their odd and even parts :
\begin{eqnarray}
\eta_{12} \, &=& \, (\eta_1(x)  +   \eta_1(-x))/2\, , \quad \quad
\eta_{11} \, = \, (\eta_1(x)  -   \eta_1(-x))/2\, ,  \\
\eta_{22} \, &=& \, (\eta_2(x)  +   \eta_2(-x))/2\, , \quad \quad
\eta_{21} \, = \, (\eta_2(x)  -   \eta_2(-x))/2\, , \quad   \quad
\phi_{12} \, = \, \cdots  \nonumber 
\end{eqnarray}
We must also decompose $\alpha(x)$ and $\beta(x)$ in odd and even
parts:
\begin{eqnarray}
\alpha_1(x)\, =\, \, {{ \alpha(x) \, - \,\alpha(-x) } \over {2}}\,, \quad
\alpha_2(x)\, =\, \, {{ \alpha(x) \, + \,\alpha(-x) } \over {2}}\,,
\quad 
\beta_1(x)\, =\, \, {{ \beta(x) \, - \,\beta(-x) } \over {2}}\,, \quad
\beta_2(x)\, =\, \, {{ \beta(x) \, + \,\beta(-x) } \over {2}}\, \nonumber
\end{eqnarray}
namely :
\begin{eqnarray}
\beta_2(x)\, &=&\, \,\,{\frac {3 \cdot {x}^{2} \cdot \left ({x}^{2}+1\right
)}{1-{x}^{2}-{x}^{4}}}\, , \quad \quad \quad
\beta_1(x)\, =\, \, {\frac {3\cdot  x}{1-{x}^{2}-{x}^{4}}}\, , \nonumber \\
\alpha_2(x)\, &=&\, \,
{\frac {3 \cdot (1+2\,{x}^{2}+2\,{x}^{4})}{\left (1-{x}^{2}\right )
\left (1-{x}^{2}-{x}^
{4}\right )}}\, , \quad \quad
\alpha_1(x)\, =\, \,{\frac {3 \cdot x \cdot \left (2+{x}^{2}+{x}^{4}\right )}
{\left (1-{x}^{2}\right )\left (1-{x}^{2}-{x}^{4}\right )}}\, , \quad
\nonumber 
\end{eqnarray}
Instead of functional relations (8.6) and (8.10) in~\cite{BoMa95},
one now has the following relations :
\begin{eqnarray}
\label{relationsalphabeta}
&&\alpha_1(x) \, -\,2 \cdot x \cdot \alpha_2(x)
\, + \,  3 \cdot x \cdot (\eta_{12}(x) \cdot \beta_2(x)\, +\,\eta_{11}(x) \cdot
\beta_1(x) )\, = \, \, 0\,, \nonumber \\
&&\alpha_2(x) \, -\,2 \cdot x \cdot \alpha_1(x)\, -\, 3\, 
\, + \,  3 \cdot x \cdot (\eta_{22}(x) \cdot \beta_1(x)\, +\,\eta_{21}(x) \cdot
\beta_2(x) )\, = \, \, 0\,, \nonumber \\
&& x \cdot \alpha_1(x) \, -\, \beta_2(x)
\, - \,  (\phi_{21}(x) \cdot \beta_1(x)\, +\,\phi_{22}(x) \cdot
\beta_2(x) )\, = \, \, 0\,, \nonumber \\
&& x \cdot \alpha_2(x) \, -\, \beta_1(x)
\, - \,  (\phi_{11}(x) \cdot \beta_2(x)\, +\,\phi_{12}(x) \cdot
\beta_1(x) )\, = \, \, 0\, 
\end{eqnarray}
where the odd and even part of the exponents generating functions
 $\, \eta_1(x)$, $\phi_1(x)$, $\eta_2(x)$,
$\phi_2(x)$, 
read  :
\begin{eqnarray}
&&\eta_{12}(x) \, = \, {\frac {{x}^{2}}{1-{x}^{4}}}\, , \qquad \quad 
\eta_{11}(x) \, = \, {\frac {x}{1-{x}^{2}}} \, , \qquad \quad 
\eta_{22}(x) \, = \, 0 \, , \qquad \quad 
\eta_{21}(x) \, = \,{\frac {{x}^{3}}{1-{x}^{4}}}  \, , \nonumber \\
&&\phi_{11}(x) \, = \,{\frac {x \cdot \left 
(2\,{x}^{2}+1\right )}{1-{x}^{4}}} \, , \qquad 
\phi_{12}(x) \, = \,2\,{\frac {{x}^{2}}{1-{x}^{2}}} \, , \qquad 
\phi_{21}(x) \, = \,{\frac {x}{1-{x}^{2}}} \, , \qquad 
\phi_{22}(x) \, = \,{\frac {{x}^{2} \cdot \left (2\,{x}^{2}+1\right
)}{1-{x}^{4}}}
 \, , \qquad \nonumber
\end{eqnarray}
Period four in the factorization scheme
(\ref{newnew}) corresponds to the occurrence of a
$\, 1-x^4\, = \, 0\, $ singularity for these exponents
generating functions.

The ``stability'' of factorization scheme (\ref{rappel}) corresponds to the
following $\, (n \rightarrow n+1)$-property : the 
exponents of the $f_n$'s occurring 
at the $m$-th step of iteration are also the one's
at $(m+1)$-th step of iteration, the $f_n$'s being
changed into  $f_{n+1}$ : at each new iteration step
one only needs to find the exponent of $f_1$ (if any).
The ``stability'' of factorization scheme (\ref{newnew}) is a straight
generalization mod.2. 
of the previous property : the exponents of the $f_n$'s occurring 
at the $m$-th step of iteration are also the one's
at $(m+2)$-th step of iteration the $f_n$'s being
changed into  $f_{n+2}$.

\subsection{Factorization scheme for $\alpha\, \ne \,0$, $\epsilon$
non generic}
Let us come back to $\, \alpha \ne 0$ with the
non-generic value $\, \epsilon \, = \, 1/2$. We consider
here  $\alpha\, = \, 396/6095\, \simeq \, .06497128$.
The factorization scheme reads :
\begin{eqnarray}
\label{newnew1/2}
&&f_1\, = \, det(M_0) \, , \quad 
M_1\,= \, K(M_0) \,, \quad
f_2\  = \, \frac{det(M_1)}{f_1} \, , \quad
M_2\,= \, \, K(M_1)  \, , \quad
f_3\, = \, \frac{det(M_2)}{f_1 \cdot f_2 } \, ,\quad 
M_3\,= \, K(M_2) \,, \nonumber \\
&&f_4\, = \, \frac{det(M_3)}{f_1^2 \cdot f_2 \cdot f_3} \, , \qquad
M_4\,= \, {{ K(M_3)} \over {f_1}} \,, \qquad 
f_5\, = \, \frac{ det(M_4)}{f_1^2 \cdot f_2^2 \cdot f_3
                \cdot f_4 } \, , \qquad 
M_5\,= \, \frac{K(M_4)}{ f_1 \cdot  f_2 
    } \, , \nonumber \\
&& f_6\, = \, \frac{det(M_5)}{f_1 \cdot f_2^2 \cdot f_3^2 \cdot f_4 \cdot 
        f_5 } \, ,\qquad 
M_6 \, = \, \frac{K(M_5)}{  f_2 \cdot f_3} \, , \, \qquad 
f_7\, = \, \frac{ det(M_6)}{f_1 \cdot f_2 \cdot 
        f_3^2 \cdot f_4^2 \cdot f_5 \cdot f_6 } \, , \qquad
 M_7 \,= \, \frac{K(M_6)}{  f_3 \cdot     f_4 }  \nonumber \\
&&  f_8\, = \, \frac{det(M_7)}{f_1^2 \cdot f_2 \cdot f_3 \cdot f_4^2 \cdot 
        f_5^2  \cdot f_6 \cdot  f_7 } \, ,\qquad  \,  
M_8 \, = \, \frac{K(M_7)}{  f_1 \cdot f_4 \cdot f_5} \, , \, \qquad 
f_9\, = \, \frac{ det(M_8)}{f_1^2 \cdot f_2^2 \cdot 
        f_3 \cdot f_4 \cdot f_5^2 \cdot f_6^2 \cdot f_7  \cdot f_8  } \, ,
\nonumber \\
&& M_9 \,= \, \frac{K(M_8)}{f_1 \cdot  f_2  \cdot  
        f_5 \cdot  f_6  }\, , \quad \quad 
 f_{10}\, = \, \frac{det(M_9)}{f_1 \cdot f_2^2 \cdot f_3^2 \cdot f_4 \cdot 
        f_5  \cdot f_6^2 \cdot  f_7^2 \cdot  f_8 \cdot  f_9  } \,
, \, \,   \,  \qquad  M_{10} \,= \, \frac{K(M_9)}{f_2 \cdot  f_3  \cdot  
        f_6 \cdot  f_7  }\, , \quad \quad  \nonumber \\
&&f_{11}\, = \, \frac{det(M_{10})}{f_1 \cdot f_2 \cdot f_3^2 \cdot f_4^2 \cdot 
        f_5  \cdot f_6 \cdot  f_7^2 \cdot  f_8^2 \cdot  f_9 \cdot f_{10} } \,
, \, \,   \,  \cdots 
\qquad M_{19} \,= \, \frac{K(M_{18})}{  f_3  \cdot  
        f_6 \cdot  f_7  \cdot  f_8 \cdot f_{11}  \cdot f_{12} 
	 \cdot f_{15}  \cdot f_{16}  }\, , \quad \quad 
\nonumber \\
&& f_{20}\, = \,  {{ det(M_{19})} \over {f_1^2 \cdot f_3 \cdot f_5^2 \cdot
f_7  \cdot
f_8^2  \cdot   f_9^2  \cdot f_{10} \cdot f_{11} \cdot 
 f_{12}^2 \cdot  f_{13}^2 \cdot  f_{14} \cdot f_{15} \cdot
f_{16}^2 \cdot f_{17}^2 \cdot f_{18} \cdot f_{19}}}  \,
, \, \,   \,   \nonumber \\
&& M_{20} \,= \, \frac{K(M_{19})}{  f_1  \cdot  
        f_5 \cdot  f_8  \cdot  f_9 \cdot f_{12}  \cdot f_{13} 
	 \cdot f_{16}  \cdot f_{17}  }\, , \quad \quad \nonumber \\
&& f_{21}\, = \,  {{ det(M_{20})} \over {f_1^2 
\cdot f_3 \cdot f_5^2 \cdot f_7 \cdot
f_8^2  \cdot f_9^2  \cdot   f_{10}^2  \cdot f_{11} \cdot f_{12} \cdot 
 f_{13}^2 \cdot  f_{14}^2 \cdot  f_{15} \cdot f_{16} \cdot
f_{17}^2 \cdot f_{18}^2 \cdot f_{19} \cdot f_{20}}}  \,
, \, \,   \,  \cdots 
\end{eqnarray}
Up to the thirteenth iteration one has the previously
described $\, (n \rightarrow n+1)$-property,
but this property is broken with $\, f_{15}$
in favor of the  $\, (n \rightarrow n+2)$-property
encountered with (\ref{newnew}).
The previously introduced odd-even-parity dependent  exponents
generating 
functions  $\eta_{ij}(x)$ and $\phi_{ij}(x)$
now read :
\begin{eqnarray}
\eta_{12}(x) \, &=& \,{x}^{2}+{x}^{6}+{x}^{10}+{x}^{12} \, , \qquad \quad 
\eta_{11}(x) \, = \, {x}^{3}+{x}^{7}+{x}^{11}
+{\frac {{x}^{15}}{1-{x}^{4}}} \, ,  \nonumber \\
\eta_{22}(x) \, &=& \, {x}^{2}+{x}^{6}+{x}^{10}
+{\frac {{x}^{14}}{1-{x}^{4}}} \, , \qquad \quad 
\eta_{21}(x) \, = \, {x}^{3}+{x}^{7}+{x}^{11}  \, , \nonumber \\
\phi_{11}(x) \,&=& \, x+2\,{x}^{3}+2\,{x}^{7}\,
+{x}^{9}+2\,{x}^{11}+{x}^{5}+2\,{x}^{13}  \, , \nonumber \\
\phi_{12}(x) \, &=& \, 
{\frac { (1+2\,{x}^{2}) \cdot x^{14}}{1-{x}^{4}}}+{x}^{2}+2\,
{x}^{4}+{x}^{6}+2\,{x}^{8}+{x}^{10}+2\,{x}^{12}\, ,   \nonumber \\
\phi_{21}(x) \, &=& \,x+2\,{x}^{3}+2\,{x}^{7}+{x}^{9}+2\,{x}^{11}+{x}^{5}
+{\frac { (1+2\,{x}^{2} ) \cdot x^{13}}{1-{x}^{4}}}\, , \nonumber \\ 
\phi_{22}(x) \, &=& \, {x}^{2}+2\,{x}^{4}+{x}^{6}+2\,
{x}^{8}+{x}^{10}+2\,{x}^{12}
 \, ,\nonumber
\end{eqnarray}
from which one deduces, from relations (\ref{relationsalphabeta}),
the rational expressions of the $\, \alpha_i$'s and   $\beta_i$'s :
\begin{eqnarray}
&&\beta_2(x)\, =\, \,\,{\frac {3 \cdot {x}^{2} \cdot 
(1\, + \, x^2 )}
{(1-x^2 )\cdot ( 1-\, x^2\, -\, x^4\, -2\, x^6\, -\,x^8\,
-2\, x^{10}\, - \, x^{12}\, -\, x^{14} )}} \, , \quad \nonumber \\
&&\beta_1(x)\, =\, \,
\,{\frac { 3 \cdot (1\, +\, x^2)\cdot (1\, +\, x^4)\cdot (1 \, + x^8  ) \cdot
x}{1\, -{x}^{2}\, -{x}^{4}\, -2\,{x}^{6}\, -{x}^{8}\, -2\,{x}^{10}\,
-{x}^{12}\, -{x}^{14}}}
\, , \nonumber \\
&&\alpha_2(x)\, =\, \, 3 \cdot {\frac
{1\, +2\,{x}^{2}\, +5\,{x}^{4}\, +4\,{x}^{6}\, +5\,{x}^{8}\, 
+4\,{x}^{10}\, +5\,{x}^{12}\, +5\,{x}^{14}\, +3\,{x}^{16}
}{(1-x^2) \cdot 
 ( 1-{x}^{2}\, -{x}^{4}\, -2\,{x}^{6}\, -{x}^{8}\, -2\,{x}^{10}\, 
-{x}^{12}-{x}^{14}
 )}}\, , \quad \nonumber \\
&&\alpha_1(x)\, =\, \, 3 \cdot x \cdot {\frac {
(2+4\,{x}^{2}+4\,{x}^{4}\, +\, 5\,{x}^{6}\, +\, 4\,{x}^{8}\, 
+5\,{x}^{10}\, +4\,{x}^{12}+\,
4\,{x}^{14} ) }{(1\, - \, x^2) 
\cdot  ( 1-{x}^{2}\, -{x}^{4}\, -2\,{x}^{6}\, -{x}^{8}\,
-2\,{x}^{10}\, -{x}^{12}\, -{x}^{14} )}}\,  
\nonumber 
\end{eqnarray}
yielding the rational expression for $\, \beta(x)$ :
\begin{eqnarray}
\label{betaunsurdeux}
&&\beta(x) \, = \, \, \, 
\,{\frac {3 \cdot x \cdot 
 (1\, +x\, +{x}^{3}\, -{x}^{16} )}{
1\, -2\,{x}^{2}\, -\, x^{6}\,+\,x^{8}\, - \,x^{10}\, +\, x^{12}\, 
+\, x^{16}}} \\
&&\qquad \, = \, \, 3 \cdot {\frac { x \cdot (1+ \, x^2 )  \cdot 
(1+x -{x}^{2} +{x}^{4}-{x}^{6}\, +x^{8}\, -{x}^{10}\, +x^{12}\, -\, x^{14} )
}{ (1-\, x^2 )\cdot 
(1\, -{x}^{2}\, -{x}^{4}\, -2\,{x}^{6}\, -{x}^{8}\, -2\,{x}^{10}\,
-{x}^{12}\, -{x}^{14} )}}\nonumber
\end{eqnarray}
The complexity growth corresponds to the (smallest) root of :
\begin{eqnarray}
\label{b16}
1\, -{x}^{2}\, -{x}^{4}\, -2\,{x}^{6}\, -{x}^{8}\, -2\,{x}^{10}\,
-{x}^{12}\, -{x}^{14} \,\, \,  = \, \, \, \, 0
\end{eqnarray}
These results have also been checked using the previously
depicted semi-numerical complexity growth evaluation 
method for $\epsilon\, = \, 1/3$ and $\alpha
\, = \,396/6095\,\simeq .06497 \cdots  \,  $.
The following value for the complexity has
 been obtained : $\lambda \, \simeq \,1.46199\, $
in good agreement with the exact algebraic value deduced from 
(\ref{b16}), namely : $\, \lambda \, \simeq \,1.46188 \cdots  $
(to be compared with the generic algebraic value of  $\, \lambda \, $,
 $\, \lambda \,\simeq \, 1.4655 \cdots  $ associated with $\,
1-x-x^3\, \, =\, 0\, $).

The singularities of (\ref{betaunsurdeux}) are in agreement with 
the dynamical zeta function calculated for these values of $\alpha$
and $\epsilon$:
\begin{eqnarray}
&&\zeta(t) \, = \, \,
{\frac
{1+t-{t}^{7}}{1-t-{t}^{2}-2\,{t}^{3}-{t}^{4}-2\,{t}^{5}-{t}^{6}-{t}^{7}}}\,
 \, = \, \, \, 
{\frac {1\, +t \cdot \left (1\, -{t}^{6}\right )}{1\, -\, 
t \cdot (1\,-t\, +t^2 )
\cdot (1\, +t\, +t^2 )^2}}
\end{eqnarray}

These calculations can also be performed, for $\alpha \ne 0$, for 
the other non-generic value of
$\epsilon $ : $\, \, \epsilon \, = \, 1/3$.
As far as the factorization scheme is concerned one gets
exactly the same scenario as the one described 
in (\ref{newnew1/2}), the breaking of the 
$\, (n \,  \rightarrow \, n+1)$-property and the 
occurrence
of a $\, (n \,  \rightarrow \, n+2)$-property
taking place with
$f_{11}$ instead of $f_{15}$ previously.
For $\, \, \epsilon \, = \, 1/3$
and, for instance, for $\alpha \, = \, 237/6095\, 
\simeq\, .038884 \cdots $, one gets the following expression
for $\beta(x) $:
\begin{eqnarray}
\label{betaunsurtrois}
&&\beta(x) \, = \, \, \,
\,{\frac {3 \cdot x \cdot \left (1+x+{x}^{3}\, -{x}^{12}\right
)}{1-2\,{x}^{2}-{x}^{6}\, +{x}^{8}+{x}^{12}}}   \\
&& \qquad \, = \, \, \,{\frac {3 \cdot x \cdot  (1\, +{x}^2 )\cdot 
 ( 1 \, + x\,  -\, x^2\, +\, x^4\, -\, x^{6}\, +\, x^8\, -\, x^{10})}{ (1-x^2 )
\cdot  (1-{x}^{2}-{x}^{4}-2\,{x}^{6}-{x}^{8}-{x}^{10} )}} \nonumber
\end{eqnarray}
Again these results have been
compared with the complexity growth
deduced from the semi-numerical method, for $\epsilon\, = \, 1/3$ and $\alpha
\, = \, 237/6095\, \simeq\, .038884 \cdots  $. We have obtained 
the following value for the complexity : $\lambda \, \simeq \,1.44865\, $
in good agreement with the exact algebraic value deduced from 
(\ref{betaunsurtrois}), namely : $\, \lambda \, \simeq \, 1.44717 \cdots $.

The singularities of (\ref{betaunsurtrois}) are in agreement with 
the dynamical zeta function calculated for these values of $\alpha$
and $\epsilon$:
\begin{eqnarray}
\zeta(t) \,  \, = \, \, {\frac
{1+t}{1-t-{t}^{2}-2\,{t}^{3}-{t}^{4}-{t}^{5}}} \, \, \,  \,  = \, \,
 \,  \, \,{\frac {1+t}{1\, -t \cdot  (1\, +\, t^2 )
\cdot  (1\, +t\, +t^2 )}}
\end{eqnarray}


\section{Choice of a initial matrix corresponding to given
values of $\epsilon$ and $\alpha$}
\label{AppendixC}
We present, in this section, a possible choice of an initial
$3 \times 3$ matrix corresponding to a prescripted value of
$\alpha$ and $\epsilon$.
From the results of Appendix \ref{AppendixA}, one has :
\begin{eqnarray}
\alpha \,\,  = \, \, 
 \frac {(x_3\,x_1\,x_2+x_2\,x_1-x_1-1)\cdot ( x_2\,x_0\,x_1+x_1\,x_0-x_0-1)}
        {(x_2-1) \cdot (x_0\,x_2-1)\cdot  x_1 }
\end{eqnarray}
and :
\begin{eqnarray}
\epsilon \,\,  = \, \, \frac {(x_1\, x_3-1)\cdot (1\, -\, x_1)\cdot
                       x_0\cdot x_2 }
                       {(1\, -\, x_2)\cdot (x_0 \, x_2-1) \cdot x_1}\, -\, 1
\end{eqnarray}
In order to perform our complexity growth calculations to
get  the factorization scheme of the transformation,
one needs to iterate a non-trivial, initial matrix as simple as possible,
in the $\, \alpha \, = \, 0\, $ case and for specific values of
 $\epsilon $ ($\epsilon \, = \, .52\, $,
 $\epsilon \, = \, 1/m\, $, ...).
Actually, let us consider a matrix of the form :
\begin{eqnarray}
M_0 \, = \, \, 
\left [\begin {array}{ccc} 1&3&x\\
\noalign{\medskip}5&2&y\\
\noalign{\medskip}-4&8&z
\end {array}\right ]
\end{eqnarray} 
The  $\, \alpha \, = \, 0\, $ condition factorizes 
as follows :
\begin{eqnarray}
\alpha \, = \, 
-\, 5\,{\frac {\left (y+5\right )\left (x+3+z\right )\left (2\,x+61-11\,y
+2\,z\right )\left (x-y-z\right )}{\left (y-5\right )^{2}\left (z+4\,x
\right )\left (2\,x-11\,y-61+2\,z\right )}}
\end{eqnarray} 
On the other hand,  expression of $\, 1\, + \, \epsilon\, $ 
is also very simple since it also factorizes :
\begin{eqnarray}
 1+\, \epsilon\, \, = \, \, 
-2\,{\frac {\left (x-z-5\right )\left (5\,x+5\,z+3\,y\right )^{2}}{
\left (y-5\right )^{2}\left (z+4\,x\right )\left (2\,x-11\,y-61+2\,z
\right )}}
\end{eqnarray} 

\section{The polynomial to find the fixed points of $k_{\epsilon}^9$}
\label{AppendixD}
The fixed points of $k_\epsilon^N$ can be found as suitable pair of roots
of two polynomials $P(z)$ and $Q(y)$. The number of pairs of
roots being relavitely small (degre($P$) $\times$ degre($Q$)), it 
is straightforward to check which are the admissible pairs.
For $\epsilon=13/25$ and $N=9$, the two polynomials
happen to verify $P(x)=Q(-x)$. We give below the expression of $P(z)$:
\begin{eqnarray}
\label{P}
P(z) \, &=& \, 
314414322376251220703125\,{z}^{18}
+1358269872665405273437500\,{z}^{17}\nonumber \\
&&
+75268905252456665039062500\,{z}^{16}
+281939167586425781250000000\,{z}^{15}\nonumber \\
&&
+4712354272080487976074218750\,{z}^{14}
+14702451771291308349609375000\,{z}^{13}\nonumber \\
&&
+115459295503780457067138671875\,{z}^{12}
+289162068299094274224609375000\,{z}^{11}\nonumber \\
&&
+235039074495145372852311328125\,{z}^{10}
-28423190864054603531819812500\,{z}^{9}\nonumber \\
&&
-129391896463704494904550698750\,{z}^{8}
-47468841855664870004702580000\,{z}^{7}\nonumber \\
&&+9768520701929861757756144700\,{z}^{6}+
8841684508557014424153308400\,{z}^{5}\nonumber \\
&&
+1497468490621088327339020023\,{z}^{4}
+77417791834794939443209320\,{x}^{3}\nonumber \\
&&
+14196266775922682562956676\,{z}^{2}
-525991376147246600507280\,z\nonumber \\
&&+4602174329226460987728
\end{eqnarray}
The actual value of $z$ for the fixed point
on $\, y \, + \, \bar{z}\, = \, 0$ is : $\, z \, \simeq \,
-.4956845\,$
$+ \, .003449852 \cdot I\, $.
Polynomials $P(z)$ and their
partners $Q(y)$ corresponding to the fixed points of $k_\epsilon^{10}$,
$k_\epsilon^{11}$, and $k_\epsilon^{12}$, 
are available in~\cite{ftpano} as well as their respective pairings of roots.


\begin{references}
\bibitem{C}
	The multi-precision library gmp is part of the GNU project.
\end{references}
\end{document}